\documentclass[a4paper,fleqn,usenatbib]{mnras}

\usepackage[T1]{fontenc}
\usepackage{ae,aecompl}

\usepackage{graphicx}	
\usepackage{caption}
\usepackage{amsmath}	
\usepackage{amssymb}	

\title[Correlating the properties of compact sources]{Correlating nonlinear properties with spectral states of RXTE data: Possible observational evidences for four different accretion modes around compact objects}

\author[Adegoke, Dhang, Mukhopadhyay, Ramadevi, Bhattacharya]{
Oluwashina Adegoke,$^{1}$\thanks{E-mail: oluwashinaa@iisc.ac.in}
Prasun Dhang,$^{1}$\thanks{E-mail: prasun@iisc.ac.in} 
Banibrata Mukhopadhyay,$^{1}$\thanks{E-mail: bm@iisc.ac.in}
\newauthor
M. C. Ramadevi,$^{2}$\thanks{E-mail: ramadevi@isac.gov.in} 
Debbijoy Bhattacharya$^{3}$\thanks{E-mail: debbijoy@gmail.com}
\\
$^{1}$Department of Physics, Indian Institute of Science, Bangalore 560012, India\\
$^{2}$Space Science Division, Space Astronomy Group, ISRO Satellite Centre, Bangalore 560017, India\\
$^{3}$Manipal Centre for Natural Sciences, Manipal Academy of Higher Education, Manipal, Udupi,
Karnataka 576104, India
}

\date{Accepted 2018, Jan 25. Received 2018, Jan 24; in original form 2017, Dec 4}

\pubyear{2018}

\begin{document}
\label{firstpage}
\pagerange{\pageref{firstpage}--\pageref{lastpage}}
\maketitle

\begin{abstract}
By analyzing the time series of \textit{RXTE/PCA} data, the nonlinear variabilities of compact sources have been repeatedly established. Depending on the variation in temporal classes, compact sources exhibit different nonlinear features. Sometimes they show low correlation/fractal dimension, but in other classes or intervals of time they exhibit stochastic nature. This could be because the accretion flow around a compact object
is a nonlinear general relativistic system involving magnetohydrodynamics. However, the more conventional way of addressing a compact source is the analysis of its spectral state. Therefore, the question arises: What is the connection of nonlinearity to the underlying spectral properties of the flow when the nonlinear properties are related to the associated transport mechanisms describing the geometry of the flow? The present work is aimed at addressing this question. Based on the connection between observed spectral and nonlinear (time series) properties of two X-ray binaries: GRS~1915+105 and Sco~X-1, we attempt to diagnose the underlying accretion modes of the sources in terms of known accretion classes, namely, Keplerian disc, slim disc, advection dominated accretion flow (ADAF) and general advective accretion flow (GAAF). We explore the possible transition of the sources from one accretion mode to others with time. We further argue that the accretion rate must play an important role in transition between these modes. 
  
\end{abstract}




\section{Introduction}

Since more than a decade, Mukhopadhyay and his collaborators (e.g. Mukhopadhyay 2004;
Misra et al. 2004; Harikrishnan et al. 2009; Karak, Dutta \& Mukhopadhyay 2010; 
and references therein) have been addressing the nonlinear properties of black hole and neutron star sources by analyzing the time series of the observed RXTE satellite data. It has been found that different temporal classes of the sources exhibit different types of nonlinear signature and fractal/correlation dimension ($D_2$: which will be defined in 
more detail in \S 2). While some of the temporal classes exhibit a low $D_2$ (which would be a possible candidate for chaos), 
others exhibit Poisson noise-like properties, namely stochastic behavior, with high $D_2$. The detailed explanation of 
the method to determine $D_2$, namely Correlation Integral (CI) method, and the related physics are discussed by
Karak, Dutta \& Mukhopadhyay (2010); Mukhopadhyay (2004); Misra et al (2004); Misra et al. (2006); Adegoke, Rakshit 
\& Mukhopadhyay (2017) etc. Although, worthy of mention is the fact that Mannattil, Gupta \& Chakraborty (2016) found no evidence of chaos or determinism in any of the GRS 1915+105 classes, the number of data points they used in each of their analyses is very small and thus makes their results highly questionable. Nevertheless, the present emphasis is not really
to identify chaos in sources, rather different $D_2$ in them and their physical meaning 
whatever be it. Even if there is no signature of chaos in them (any small $D_2$ is not the signature of chaos), the different 
$D_2$ among them is apparent which can not be neglected and something differentiable must be there between the respective 
time series. On the other hand, the compact sources, particularly black holes (e.g. GRS~1915+105), are shown to be classified into $3-4$ spectral states (Migliari \& Belloni 2003; Reig, Belloni \& van der Klis 2003). Analyzing spectral state is a powerful tool to understanding the geometry of the flow, whether it is a Keplerian disc (Shakura \& Sunyaev 1973) dominated flow or a sub-Keplerian/corona (Narayan \& Yi 1995; Rajesh \& Mukhopadhyay 2010) dominated flow. Hence, obviously the question arises, do the observed spectral states link with the nonlinear nature of the flow?

Indeed, Miyamoto et al. (1992), van der Klis (1995), Tanaka \& Shibazaki (1996), Liang (1998) argued that black holes exhibit different spectral states. They can be classified mainly into three canonical states: low/hard (Tananbaum et al. 1972), intermediate
(Belloni et al. 1997; M\'endez \& van der Klis 1997) and high/soft (Tananbaum et al. 1972). For example, despite the fact that the black hole source GRS~1915+105 has been shown to exhibit twelve different temporal classes (e.g. Belloni et al. 2000),
its variability can be understood as the transitions only between three basic spectral states (Klein-Wolt et al. 2002; Migliari \& Belloni 2003; Reig, Belloni \& van der Klis 2003), namely $A, B$ and $C$. While the first two states exhibit a soft spectrum with $B$ having a higher count-rate and a harder spectrum, state $C$ corresponds to a low count-rate and a hard spectrum. Therefore, in general, the X-ray spectrum of the source can be 
modelled as a combination of a soft disc blackbody representing the radiation from the Keplerian accretion disc and a hard power law extending to $\lesssim100$ keV which corresponds to a Comptonizing sub-Keplerian flow. Reig, Belloni \& van der Klis (2003), in order to correlate disc dynamics with spectral states for GRS~1915+105, interpreted
the rapid X-ray variability involved with the states $A$ and $B$ as filling of the inner 
region of an accretion disc and the state $C$ as disappearing of the disc, due to the onset of a thermal/viscous instability. However, the same source, based on radio observations with the VLA, is also shown to exhibit collimated relativistic jet (Mirabel \& Rodr\'iguez 1994). Pooley \& Fender (1997) suggested a possible correlation between X-ray dips observed in the highly complex $\beta$ class of GRS~1915+105 and its radio flares. This temporal class, in a certain observation (Migliari \& Belloni 2003), 
alone itself exhibits all three spectral states. On the other hand, the radio flare is shown to correspond to the state $C$ (Klein-Wolt et al. 2002). This implies the connection of radio jets with the disc instability, which might correspond to the variations of local mass accretion rate and mass loss. Remillard \& McClintock (2006) also classified the source into three different spectral states, namely thermal, hard and steep power law, based on the contributions from the model disc multicolor blackbody, power law and appearance of QPO. Furthermore, Fender \& Belloni (2004) outlined the observational properties of GRS 1915+105, comparing it with other sources with the aim of constructing a simple model for the disc-jet coupling. Through a detailed analysis of the states of GRS 1915+105 vis-a-vis the canonical states of other black hole candidates, they argued that its properties are not very dissimilar from the X-ray states observed in other black hole binaries.

While several of the above facts reveal GRS~1915+105 to be a canonical black hole, many of them indicate the source to be peculiar with respect to other sources. Other compact sources (including neutron stars which, although reveal differences compared to black holes 
due to the presence of a surface), e.g. Sco~X-1, Cyg~X-1, Cyg~X-2, Cyg~X-3, are also expected to exhibit some (if not all) of the canonical spectral behavior(s) as seen in GRS~1915+105.

Interestingly, on the basis of extensive X-ray timing analysis, 
sources Sco~X-1, Cyg~X-2 and some temporal classes of GRS~1915+105 seemed to exhibit chaos or at least of low $D_2$ (fractal), while Cyg~X-1, Cyg~X-3 and some other temporal classes of GRS~1915+105 appeared 
to be stochastic with high, unsaturated $D_2$ earlier (see, e.g., Mukhopadhyay 2004; Misra et al. 2004; Karak, Dutta \& Mukhopadhyay 2010). Note that while the nature of Cyg~X-3 is still uncertain, an attempt was made to identify it based on nonlinear time series analysis by Karak, Dutta \& Mukhopadhyay (2010) in which they argued it to be possibly a black hole. However, nonlinear/temporal behaviors are generally expected to be related with spectral states. It is unlikely that the spectral nature appears randomly as against the underlying 
nonlinear properties. In the present paper, we aim at finding a possible relation between spectral states and temporal classes for compact sources: one black hole and one neutron star to start with, from the earlier analyzed systems, combining the results obtained from spectral and timing analyses independently. We call the relationship between the time series behavior and the spectral behavior "correlation".

Although previous works considered the evolution of the spectral and timing properties 
of accreting sources (e.g., Homan \& Belloni 2005; Ingram et al. 2011; Kalamkar et al. 2015), none of them applied the CI technique and aimed at determining the correlation of nonlinear nature with spectral state of the flows, which would help in modeling the flow physics. As the spectral states reveal the dynamical nature of the flow, whether it 
is a multi-color blackbody (diskbb model in XSPEC) dominated Keplerian disc or power law (PL) dominated corona/sub-Keplerian disc, this in turn will help in correlating nonlinearity with the fundamental processes in the flow. We aim to understand the above mentioned correlation in terms of various model classes of accretion flow, particularly those that are already well established, namely: Keplerian disc (Shakura \& Sunyaev 1973; Novikov \& Thorne 1973), advection dominated accretion flow (ADAF; Narayan \& Yi 1994), general advective accretion flow (GAAF; Rajesh \& Mukhopadhyay 2010) and slim disc (Abramowicz et al. 1988). We propose a cycle of flows converting among the accretion modes (i.e aforementioned accretion flows) with the change of a fundamental parameter describing the system. While we already have some idea about model classes (modes) of the flows and their conversion among each other, there are uncertainties in the modelling. For example, what is the actual number of independent variables and equations describing GAAF and ADAF? Is the magnetic field playing the key role in determining the dynamics explicitly? If so, are all the field components equally important? How exactly is the cooling effect determining the flow dynamics and how efficient are the various components of the cooling or radiation mechanisms? All of these important questions determine the number of variables, which finally control the flows, and then the number of equations, which determine the flow structure. Therefore, our venture to obtain the said correlation from observed data will enormously help in correctly modeling the various classes of accretion flows. None of the previous studies, which investigated simultaneously the temporal and spectral behaviors of the sources, could enlighten on this issue. Our study 
unveils the issue related to the fundamental modeling of complex accretion flows,
based on observed data by using both spectral and nonlinear timing analysis methods to deduce the actual degrees of freedom determining the flow dynamics.

In the following section (\S 2), we briefly outline the CI method to quantify the nonlinearity of sources. Next 
in section 3, we analyze the spectral states of the sources along with the underlying nonlinear nature. In this context, first we analyze the twelve RXTE Observational IDs (ObsIDs) representing each of the twelve temporal classes of the black hole GRS~1915+105 in section 3.1.
Subsequently, we concentrate on a neutron star Sco~X-1 in section 3.2. More so, we perform similar analysis 
on simulated data representing a model accretion flow, discussed in section 3.3. Finally we conclude in the 
last section (\S 4) with discussions.

\section{Correlation Integral method}

To reconstruct the dynamics of a nonlinear system from a time series, the delay embedding technique of Grassberger \& Procaccia 
(1983) is one of the standard methods used. It requires constructing a CI which is the probability that two arbitrary points in 
phase space are closer together than a distance $r$. Because the number of governing equations or variables in the system is not 
known a priori, one is required to construct the dynamics for different embedding dimensions $M$, which is the number of points 
constructed from a time series. In using this method, points (vectors) of length (dimension) $n$ are produced from the 
time series $s(t_{i})$ by applying a time delay $\tau$ such that
\begin{center} 
$x(t_{1})=s(t_{1}), s(t_{2}),...,s(t_{n})$,\\
$x(t_{2})=s(t_{1}+\tau), s(t_{2}+\tau),...,s(t_{n}+\tau)$,\\
.\\
.\\
.\\
$x(t_{M})=s(t_{1}+(M-1)\tau), s(t_{2}+(M-1)\tau),...,s(t_{n}+(M-1)\tau)$.
\end{center}
A necessary condition is that the times $t_{i}$ be equally spaced and should be chosen such that the number of data 
points is sufficiently large. For instance, time bin of $0.1\,\mathrm{s}$ has been chosen for the \textit{RXTE} data analyzed in this paper. 
Furthermore, $\tau$ has to be chosen in a way that the vectors are not correlated i.e. when autocorrelation function of 
$x(t_{i})$ reaches its first minima or when it approaches zero.

A quantitative picture is provided by the correlation dimension $D_{2}$. Computationally, it involves choosing a 
large number of points in the reconstructed dynamics as centers and the number of points that are within a distance $r$ 
from the center averaged over all the centers is defined to be the correlation integral $C_{M}(r)$ written as
\begin{eqnarray}
C_{M}(r)=\frac{1}{N({N_{c}-1})}\mathop{\sum^{N}\sum^{N_{c}}}_{i=1\ j\neq i\ j=1}{H(r-|{x_{i}-x_{j}}|)},
\end{eqnarray}
where $N$ and $N_{c}$ are respectively the number of points and the number of centers, $x_{i}$ and $x_{j}$ are the 
reconstructed vectors while $H$ is the Heaviside step function. The correlation dimension $D_{2}$ being just a 
scaling index of the variation of $C_{M}(r)$ with $r$ and is expressed as
\begin{eqnarray}
D_{2}=\mathop {\lim }\limits_{r \to 0}\left(\frac{d\rm{log}C_{M}(r)}{d\rm{logr}}\right).
\end{eqnarray}
In principle, the value of $D_{2}$ can help to determine the effective number of differential equations describing 
the dynamics of the system. The linear part of the plot of $\rm{log}[C_M(r)]$ against $\rm{log}[r]$ can be used to estimate $D_{2}$ whose value depends on $M$. The nonlinear dynamical properties of the system is revealed from the variation of $D_{2}$ with $M$. If for all $M$, $D_{2}\approx M$ then the system is stochastic. On the other hand, the system 
is deterministic if initially $D_{2}$ increases linearly with $M$ until it reaches a certain value and saturates. 
This saturated value of $D_{2}$ is then taken as the fractal/correlation dimension of the system, plausibly a signature of chaos (subject to the confirmation by other methods).
In simple terms, a system is chaotic if any two consecutive trajectories of the system, while divergent from each other, are related by some law (although instantaneously, it may appear random), e.g. underwater current in the ocean in the abysmal zone. On the other hand, a system is stochastic or random if any two consecutive trajectories of the system 
are not related by any physics, Poisson noise is an example. A chaotic or deterministic system generally exhibits self-similarity, a property not present in stochastic systems. Therefore the nonlinear times series analysis technique briefly described above could be a robust method for studying the dynamics of an accretion system from its lightcurve.

\section{The correspondence between spectral states and nonlinear properties} 

We begin our analysis with GRS~1915+105, which already has been investigated in detail in other contexts. Therefore, it will be easier to make connection between our results and the ones already known for this source. For this purpose, we consider the same ObsIDs which were considered in earlier or previous works (e.g. Belloni et al. 2000;
Mukhopadhyay 2004). Once our method and results are established based on this source,
we then apply our method to Sco X-1. We consider the energy range $3-25$ keV, for our \textit{RXTE} data (spectral) analysis.

\subsection{GRS~1915+105}

\subsubsection{General features obtained in time series}

The time series of GRS~1915+105 for twelve different ObsIDs corresponding to each of the twelve temporal classes were already studied to determine the underlying nonlinearity
(Mukhopadhyay 2004; Misra et al. 2004). Four of the temporal classes are shown to have low $D_2$, namely fractal, referred to as F (may be chaos) and three of them to be stochastic with unsaturated $D_2$, referred to as S. While Mukhopadhyay (2004) showed the remaining five to be semi/non-stochastic (NS), which have large $D_2$ but deviated from ideal stochastic feature, Misra et al. (2006) argued, based on singular-value-decomposition technique, that three of these five classes are actually chaotic, one is limit cycle (LC) and the remaining one is stochastic. Even if it is yet to be confirmed by other methods, whether any of the twelve classes is chaotic or not, there is a clear evidence that different temporal classes have varied nonlinear properties, as reflected in the corresponding $D_2$. What about the corresponding spectral states? 

\begin{table*}
\begin{center}
\small
\caption{GRS~1915+105: Basic flow classes are highlighted keeping within triple-lines -- see Table \ref{tab:tab3} in Appendix for best fit parameters}
\label{tab:tab1}
\begin{tabular}{cccccccccccccccccccccc}\\
\hline
\hline
ObsID & class     &  behavior & diskbb & PL & GA &  SI  &
$\chi^{2}/\nu$ & state & count-rate  & $L$ \\
\hline
\hline
10408-01-10-00 & $\beta$  & F & $46$ & $52$ & $2$ & $3.25^{+0.03}_{-0.03}$ & $0.9683$ & D-P & $3726$ & $0.3901$\\
\hline
20402-01-37-01 & $\lambda$ &  F & $54$ & $46$ & $-$ & $2.96^{+0.03}_{-0.03}$ & $0.9532$ & D-P & $2902$ & $0.3119$\\
\hline
20402-01-33-00 & $\kappa$ & F & $49$  & $51$ & $-$ & $2.99^{+0.02}_{-0.02}$ & $1.1740$ & D-P & $2526$ & $0.2720$\\
\hline
\hline
\hline
10408-01-08-00 & $\mu$ &  F & $56$ & $41$ & $03$ & $3.26^{+0.04}_{-0.04}$ & $0.9603$ & DD & $3891$ & $0.4040$\\
\hline
\hline
\hline
20402-01-45-02 & $\theta$ &  F & $11$ & $88$ & $01$ & $3.10^{+0.02}_{-0.02}$ & $1.2880$ & PD & $3202$ & $0.3536$  \\
\hline
10408-01-40-00 & $\nu$ &  F & $28$ & $72$ & $-$ & $2.93^{+0.02}_{-0.02}$ & $1.3570$ & PD & $3202$ & $0.3498$\\
\hline
\hline
\hline
20187-02-01-00 & $\alpha$ &  F & $23$ & $77$ & $-$ & $2.39^{+0.02}_{-0.02}$ & $0.5198$ & PD & $1136$ & $0.1341$ \\
\hline
\hline
\hline
20402-01-03-00 & $\rho$ &  LC & $28$ & $72$ & $-$ & $2.86^{+0.02}_{-0.02}$ & $0.7870$ & PD & $2431$ & $0.2659$\\
\hline
10408-01-17-00 & $\delta$ &  S & $48$ & $50$ & $02$ & $3.43^{+0.03}_{-0.03}$ & $1.4870$ & D-P & $2601$ & $0.2683$\\
\hline
10408-01-12-00 & $\phi$ & S & $50$ & $34$ & $16$ & $3.84^{+0.03*}_{-0.03}$ & $1.1080$ & DD & $2006$ & $0.1995$\\
\hline
\hline
\hline
20402-01-56-00 & $\gamma$ & S & $60$ & $31$ & $09$ & $3.78^{+0.04}_{-0.04}$ & $1.1200$ & DD & $3521$ & $0.3598$\\
\hline
\hline
\hline
10408-01-22-00 & $\chi$ & S & $09$ & $89$ & $02$ & $3.00^{+0.01}_{-0.01}$ & $1.0700$ & PD/LH & $1862$ & $0.2107$\\
\hline
\hline
\hline
\end{tabular}
\end{center}
{Columns:- 
1: RXTE observational identification number (ObsID).
2: Temporal class.
3: The behavior of the system (F: low correlation/fractal dimension;
S: Poisson noise like stochastic; LC: limit cycle).
4: $\%$ of multi-color blackbody component.
5: $\%$ of power law component.
6: $\%$ Gaussian line component (XSPEC model gauss).
7: Powerlaw photon spectral index.
8: Reduced $\chi^2$.
9: Spectral state (DD: disc dominated; D-P: disc-power law contributed; PD: 
power law dominated; LH: low hard).
10: Model predicted count-rate (cts/s).
11: Total luminosity in $3-25$ keV in units of Eddington luminosity
for the black hole mass $14M_\odot$ (Greiner et al. 2001).
N.B: Parameters with '$*$' are kept fixed at their fiducial values for acceptable fits.
}
\end{table*}

\subsubsection{Spectral features in F cases}

\begin{figure*}
\centering
 \begin{tabular}{@{}cc@{}}
  \includegraphics[angle=-90,width=.50\textwidth]{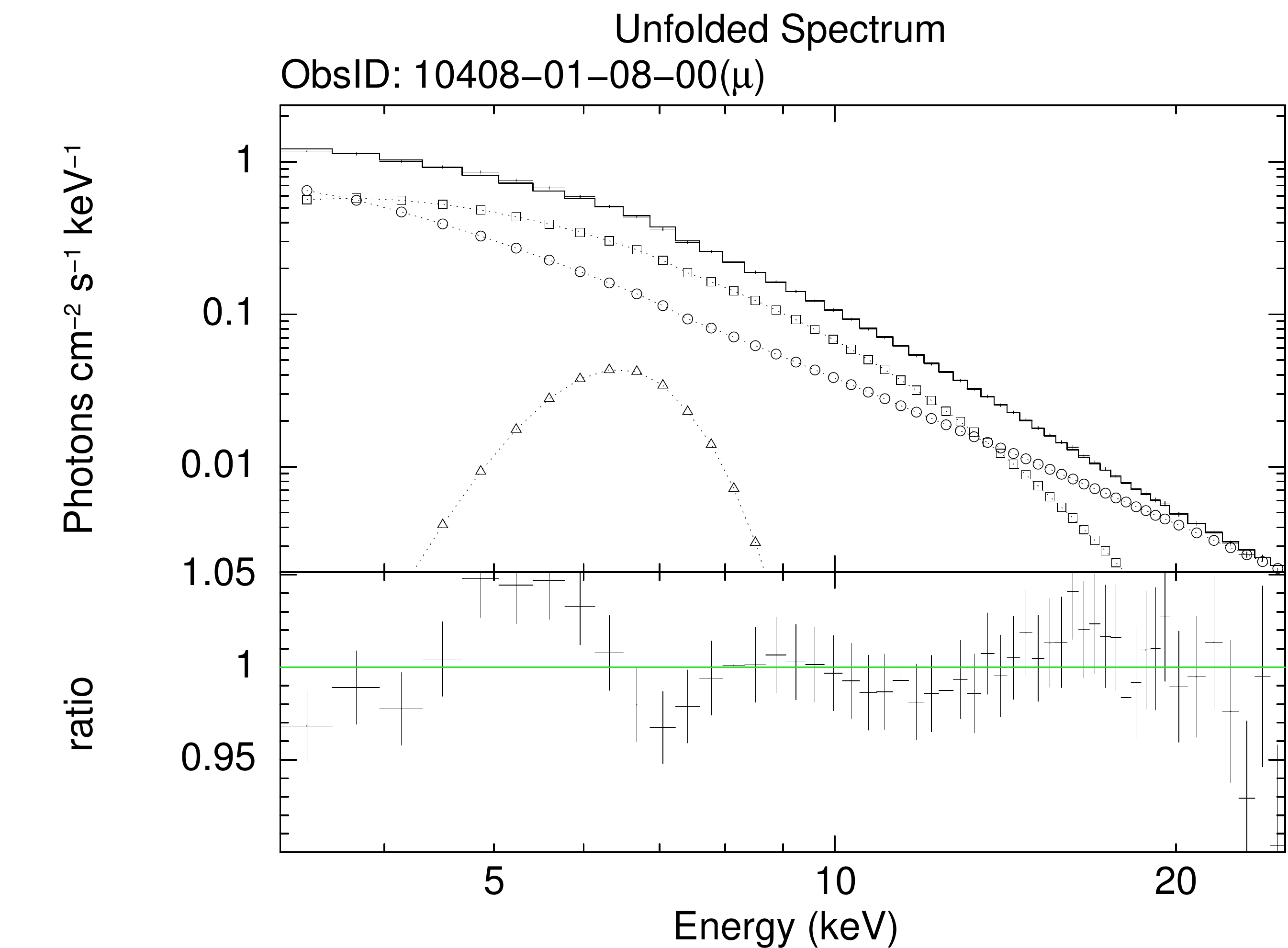} &
  \includegraphics[angle=-90,width=.50\textwidth]{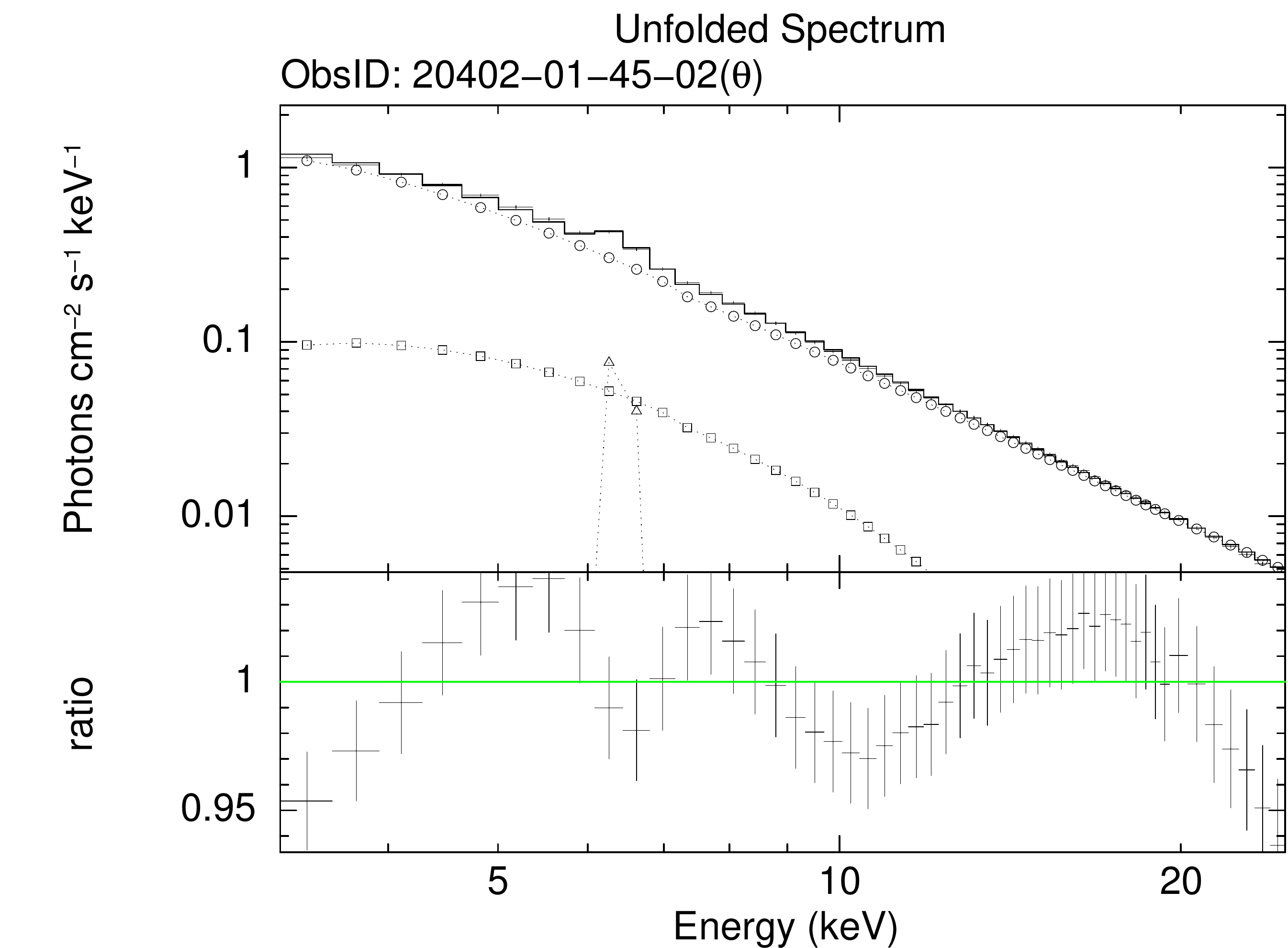} \\
  \\
  \includegraphics[width=.52\textwidth]{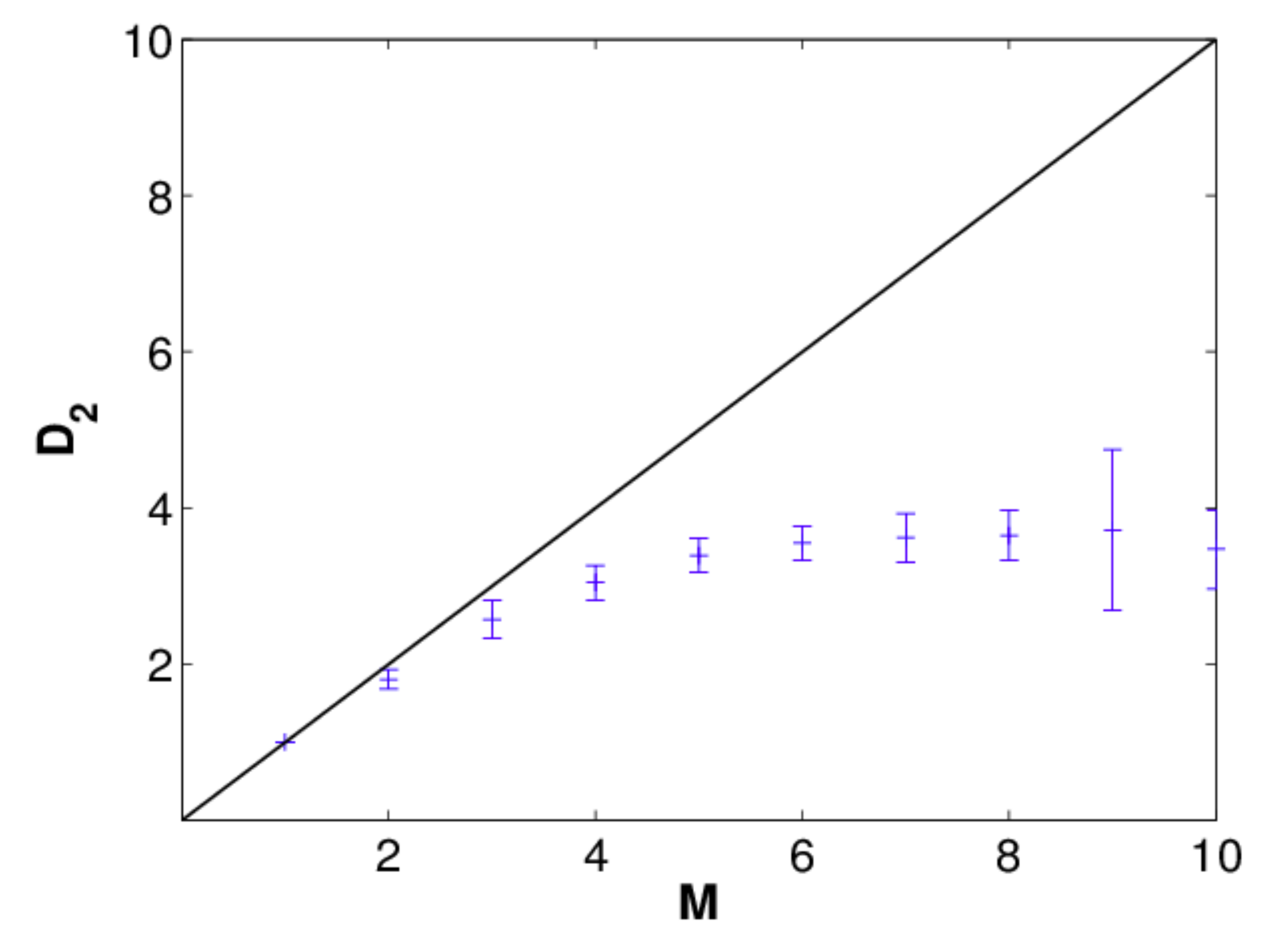} &
  \includegraphics[width=.52\textwidth]{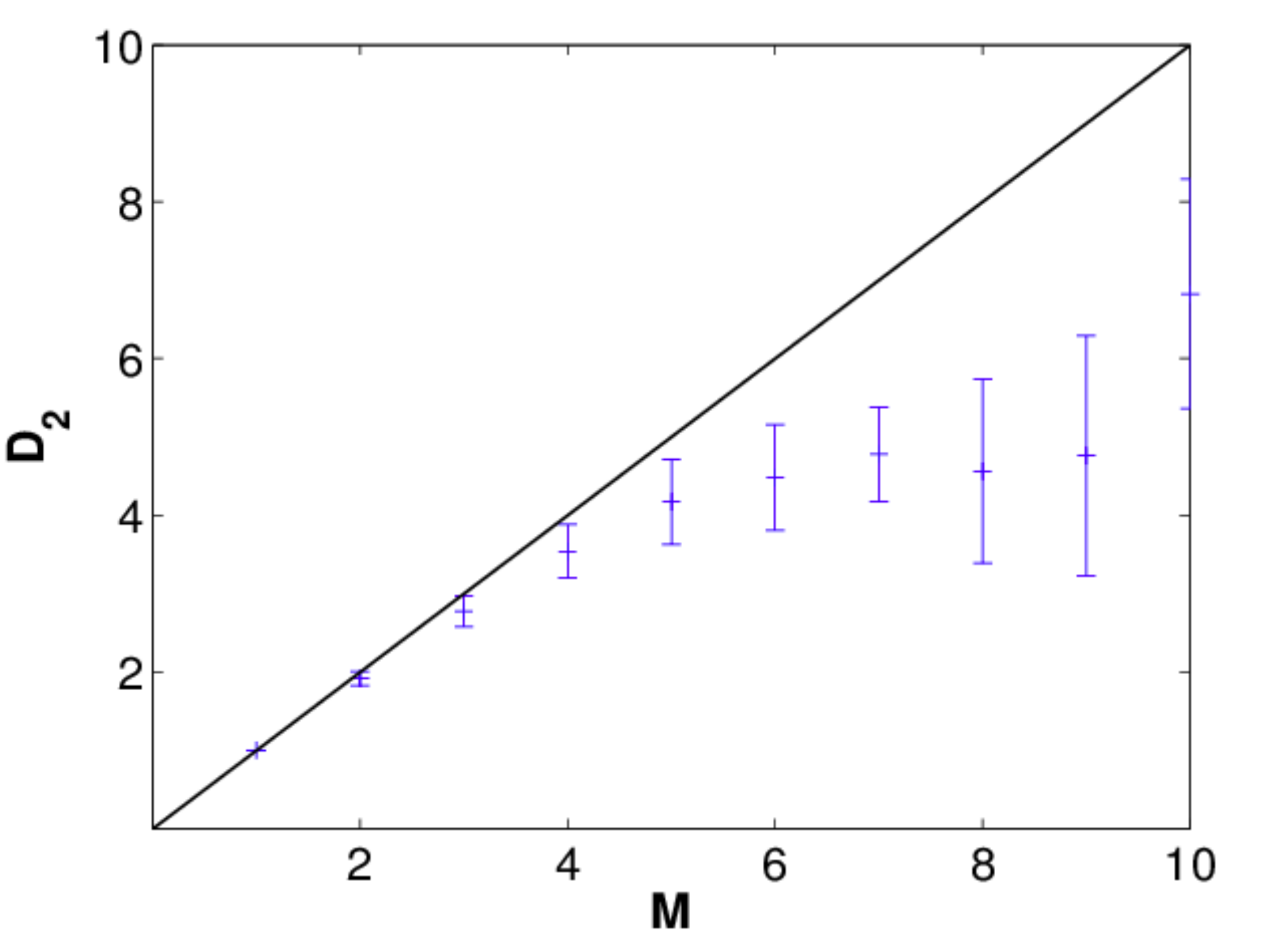} \\
 \end{tabular}
\caption{Top panels: Unfolded spectra for $\mu$ class (top-left) and $\theta$ class (top-right) of GRS~1915+105. The curves with square, circle and triangle symbols represent respectively diskbb, power law and Gaussian (iron) line contribution and the topmost solid curve in each panel represents the overall spectrum. 
Bottom panels: Variation of correlation dimension as a function of embedding dimension with errors for $\mu$ class (bottom-left) and $\theta$ class (bottom-right) of GRS~1915+105. The solid diagonal line in each panel represents the ideal stochastic curve. See Table \ref{tab:tab1} for other details.
}
\label{fig:fig1}
\end{figure*}

Table \ref{tab:tab1} shows that out of seven classes with low saturated values of $D_2$, indicated by F (based on Misra et al. 2006), one of them ($\mu$) exhibits diskbb dominated (DD) spectrum. We consider a state to be DD if the ratio of contributions from diskbb to PL is $\gtrsim 4/3$ and powerlaw dominated (PD) if the ratio of contributions from PL to diskbb is $\gtrsim 4/3$. However, three of them ($\beta, \lambda, \kappa$) reveal almost equal contributions from diskbb and power law, named as D-P spectra. But all four ($\beta, \lambda, \kappa, \mu$) of them exhibit high to very high (for DD class $\mu$) count-rate (and hence luminosity) with steep power law photon spectral indices (SIs) $\sim 3.25, 2.96, 2.99, 3.26$ respectively. Figure \ref{fig:fig1} depicts the representative cases of DD and PD spectra having F in time series. In Fig. \ref{fig:fig1} (top-left), we show unfolded energy spectra for class $\mu$, describing its DD nature. Note that as significant PL contributes to the energy, the class could not reveal a pure soft state, but reveals a very high (VH) state (e.g. Zdziarski \& Gierlinski 2004). We also show the variation of $D_2$ with $M$ (Mukhopadhyay 2004; Misra et al. 2004) saturating to a low $D_2\sim 4$ for the time series of same class (bottom-left).
All the remaining F classes ($\theta, \nu, \alpha$) exhibit PL dominated spectra, with 
diskbb contribution as low as $\sim 11\%$ and the PL as high as $\sim 88\%$ for $\theta$, and $\alpha$ showing the lowest luminosity among the classes,
revealing PD harder spectral states, as given in Table \ref{tab:tab1}. The class $\alpha$ is hardest, among all the classes, with the SI $\sim 2.39$ and very low count-rate. Other classes ($\theta$ and $\nu$) exhibit reasonably higher count-rate, which may partly be due to their non-negligible diskbb components. Note that it was already reported that the class $\theta$ is a peculiar class (e.g. Belloni et al. 2006) differing quite a bit from the features exhibited by other canonical black holes. According to Belloni et al. (2006), it is a combination of the spectral states $A$ and $C$ having a high frequency QPO. Hence, it exhibits a high flux/luminosity with PL dominance. Figure \ref{fig:fig1} (top-right) also demonstrates that the class $\theta$ exhibits PL dominance in the unfolded energy spectra, while its time series clearly reveals a saturated $D_2$ (bottom-right).

\subsubsection{Spectral features in S cases}

\begin{figure*}
\centering
 \begin{tabular}{@{}cc@{}}
  \includegraphics[angle=-90,width=.50\textwidth]{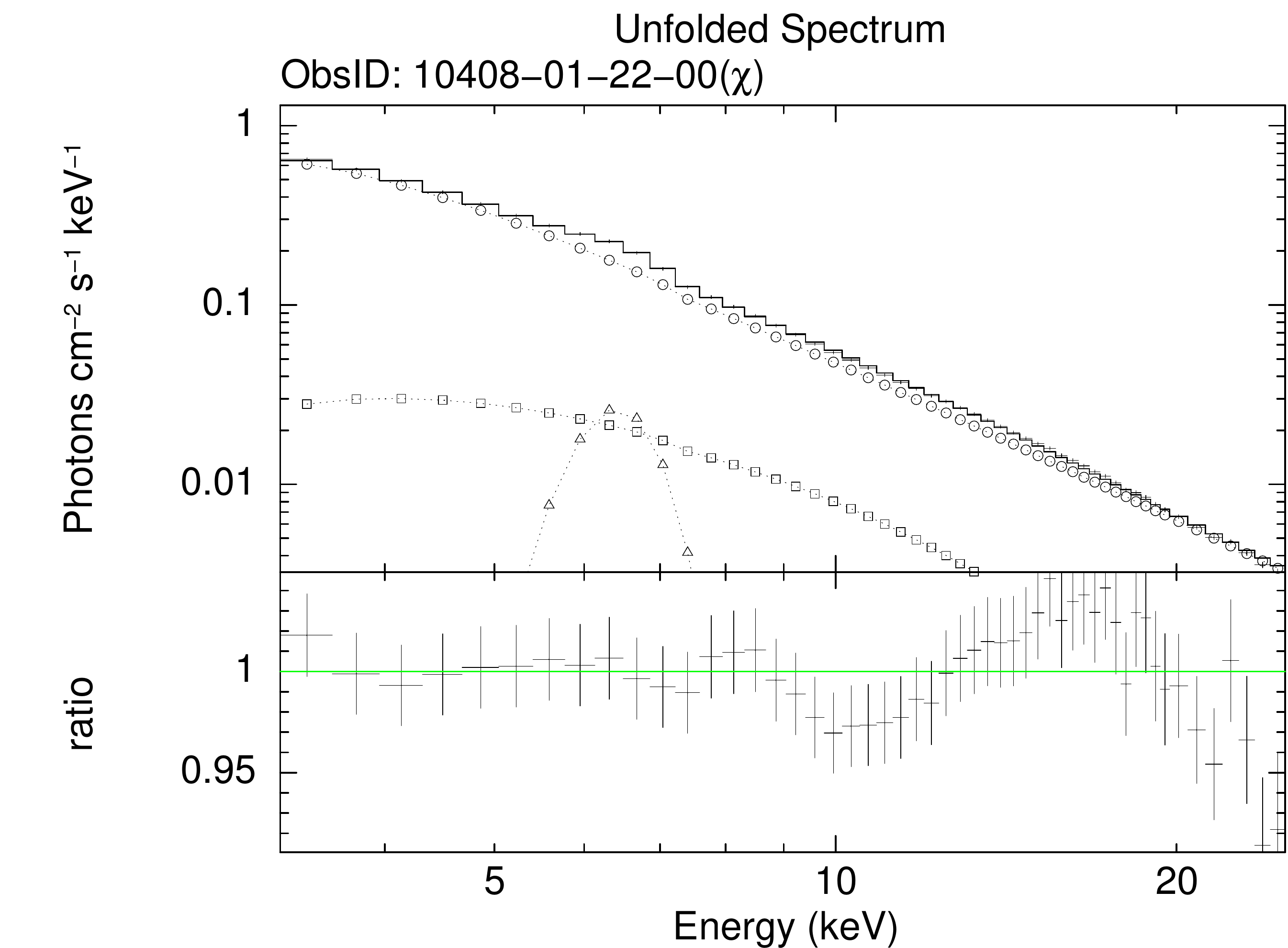} &
  \includegraphics[angle=-90,width=.50\textwidth]{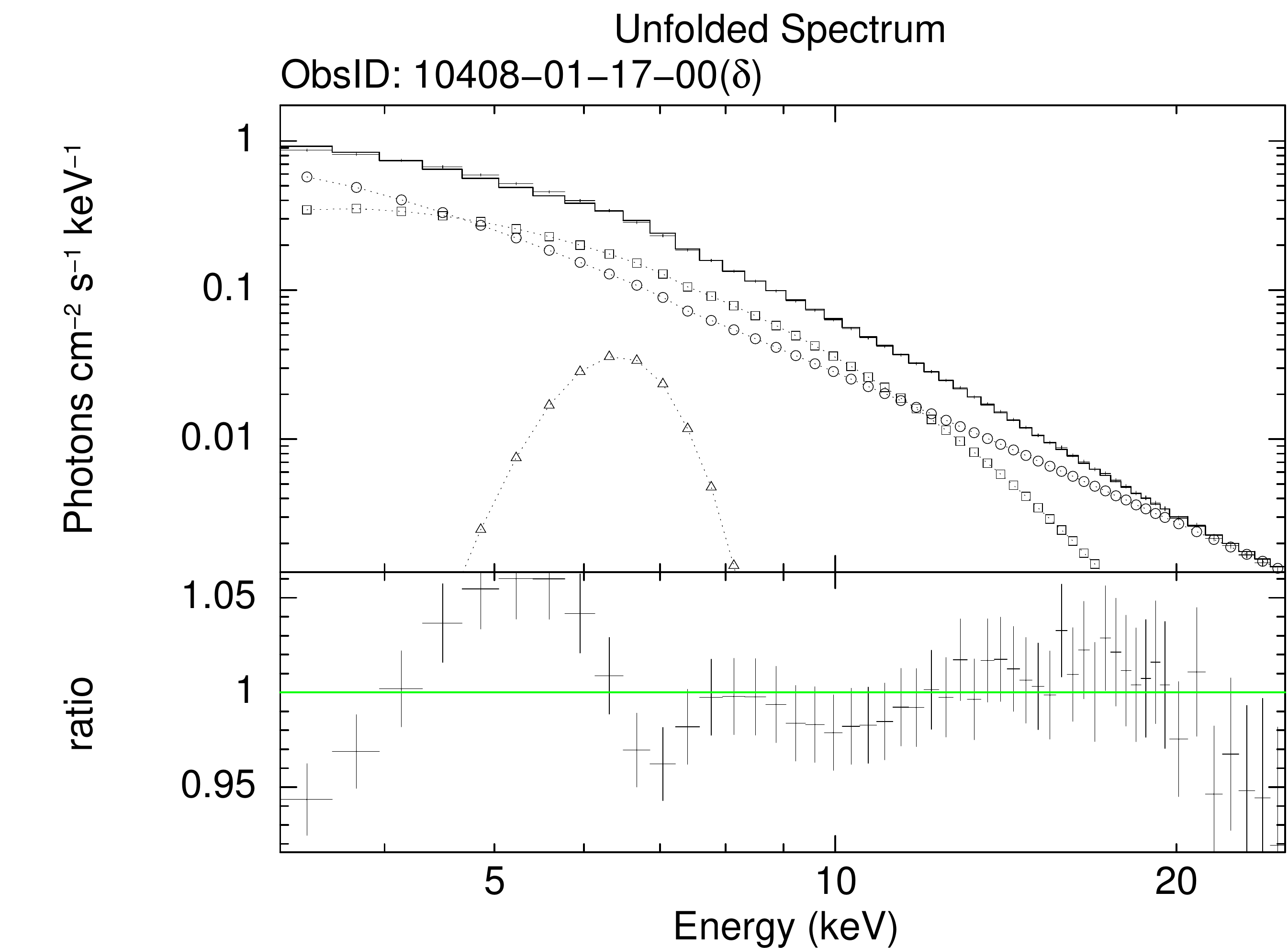} \\
  \\
  \includegraphics[angle=360,width=.52\textwidth]{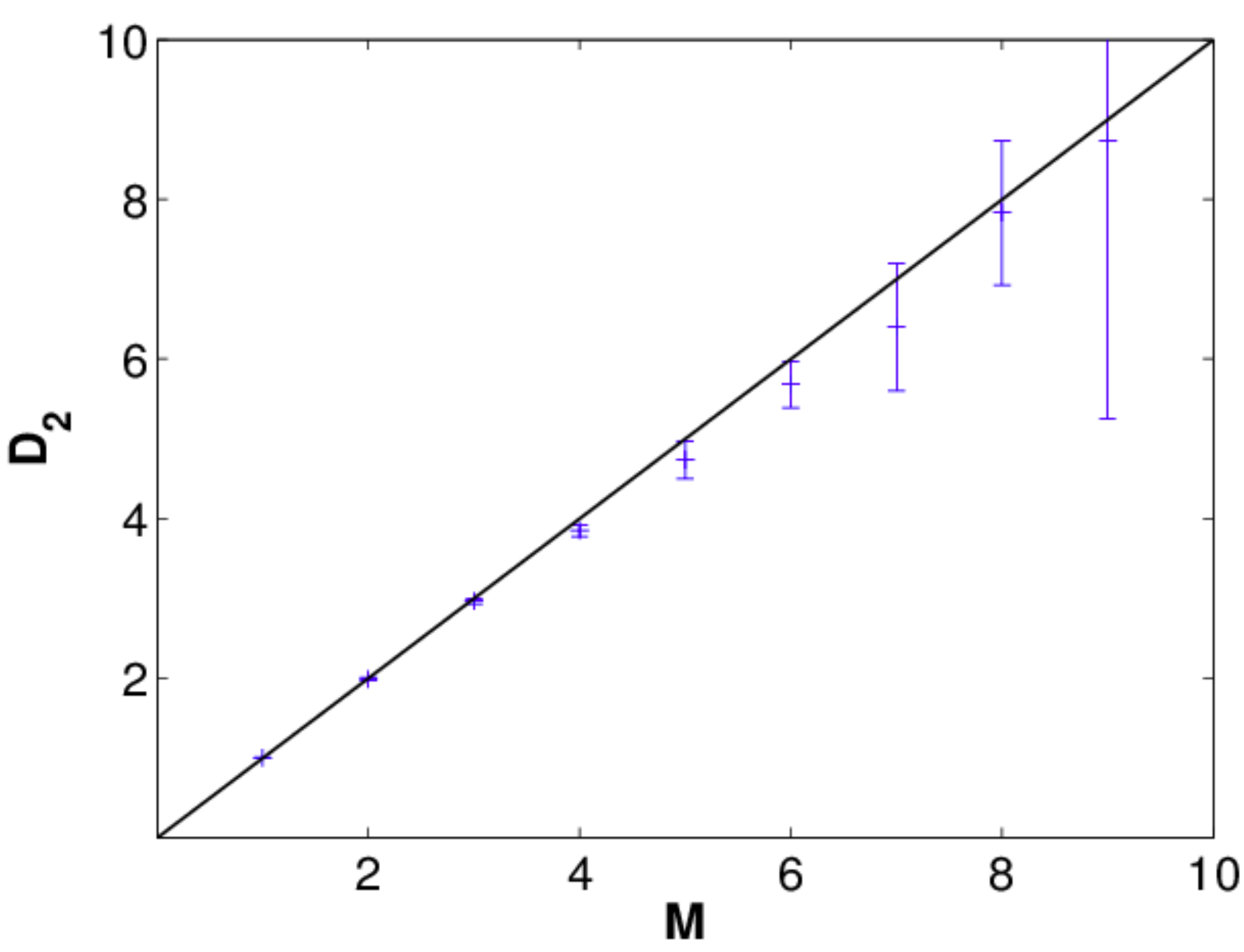} &
  \includegraphics[angle=360,width=.52\textwidth]{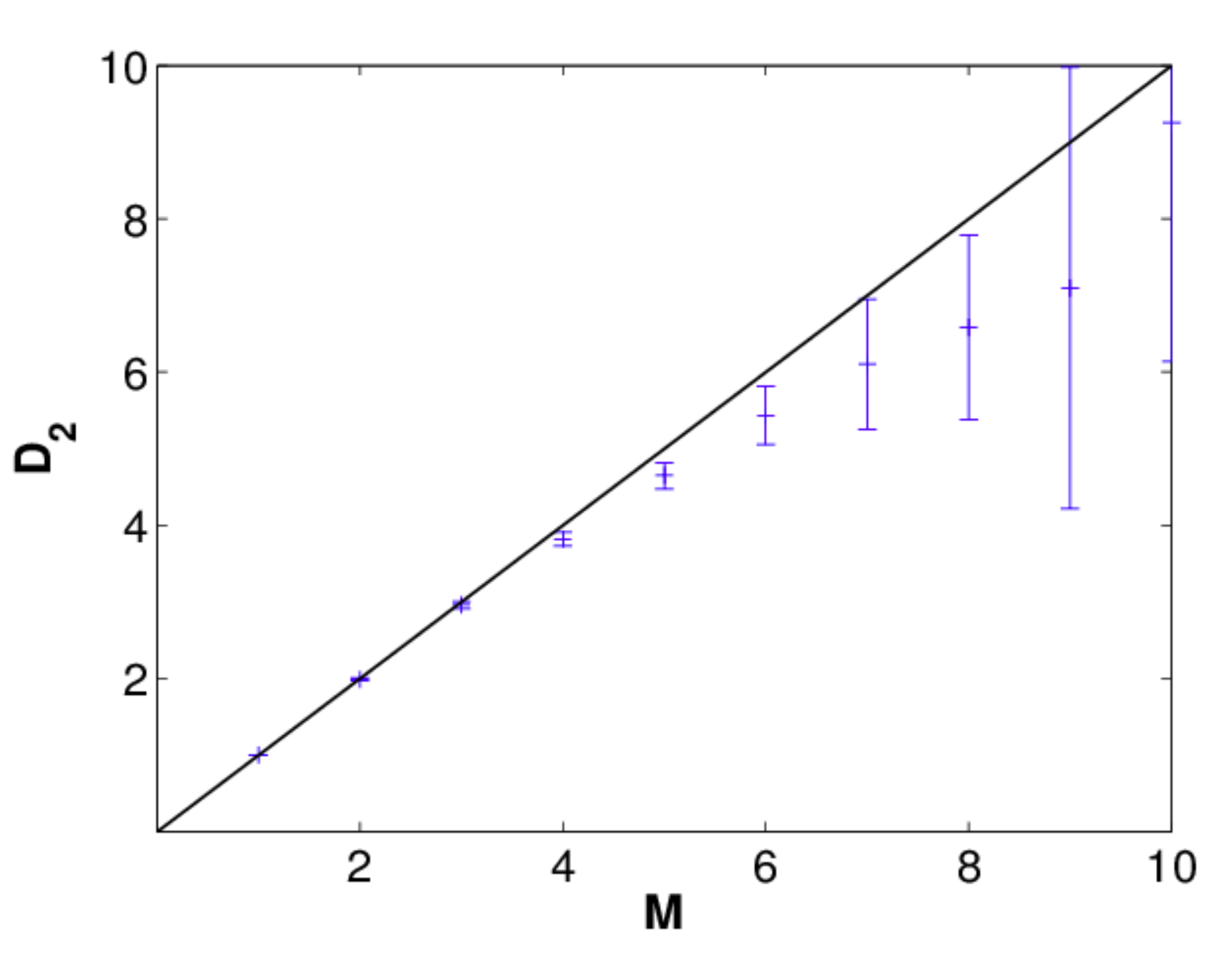} \\
 \end{tabular}
\caption{Same as in Fig. \ref{fig:fig1}, except the results for classes $\chi$ (top-left, bottom-left) and
$\delta$ (top-right, bottom-right) are shown.
}
\label{fig:fig2}
\end{figure*}

Of the four classes showing stochastic nature, denoted by S in Table \ref{tab:tab1}, two of them ($\gamma, \phi$) are in DD states with at least $50\%$ diskbb contribution and one ($\delta$) is in D-P state, with all having steep SIs $\gtrsim 3.0$. While $\delta$ and $\phi$ exhibit lower count-rates and flux/luminosity than most DD and D-P classes showing 
F, $\gamma$ class shows high flux/luminosity. This is understood from its highest diskbb content, compared to the other ones (as evident from Table \ref{tab:tab1}). Interestingly, while the classes $\gamma$ and $\phi$ both are DD, the latter exhibits almost half the flux of the former which will be discussed in \S3.1.4. Figure \ref{fig:fig2} depicts the representative cases of DD/D-P and PD spectra having S in time series. In Fig. \ref{fig:fig2} (right panels), we describe the combination of temporal and spectral behaviors for
the class $\delta$ (exhibiting a typical combination of S and D-P). The fourth class ($\chi$) showing S, however, is very special, whose spectrum behaves similar to that of the hard state of Cyg~X-1, a canonical black hole source, exhibiting a low/hard PD state (see Fig. \ref{fig:fig2} top-left). Although it exhibits a powerlaw tail, it also contains a non-negligible diskbb component (confirmed by f-test statistics) and steeper overall SI compared to the class $\alpha$ showing the lowest count-rate. However, compared to most of the classes, it exhibits an overall harder SI, justifying its lower count-rate/flux. It also should be noted that the $\chi$ class is sub-divided into $\chi_1$, $\chi_2$, $\chi_3$ and $\chi_4$ (Belloni 2000) and SIs of the $\chi_2$ and $\chi_4$ classes are harder (flatter) than those of $\chi_1$ and $\chi_3$. Since, we consider the $\chi_1$ class, its SI appears to be higher. This further makes its flux non-negligible compared to other classes. Its time series behavior is shown in Fig. \ref{fig:fig2} (bottom-left). The remaining class $\rho$ exhibits a combination of PD spectral state and non-Poissonian dynamics in time series (see Misra et al. 2006), making it similar to the classes $\theta, \nu, \alpha$. Note that the classes showing DD spectrum contain nonzero PL components and those showing PD also contain nonzero diskbb components. Hence none of them exhibit pure high/soft (HS) and low/hard (LH) states respectively. In fact, some classes represent (repeated) conversion from DD to PD (e.g. $\lambda$) and hence overall exhibit D-P states or non-pure states.

\subsubsection{Identifying accretion modes as the combination of observed spectral states (HS/LH) and nonlinear classes (F/S)}

There are two classes of nonlinear temporal behavior: F and S, and two extreme spectral states: HS (diskbb) and LH (PL), although, in general, the overall spectral states in different flows happen to be DD, PD, D-P. Hence, there are four extreme possible combinations which may arise in flows: F and HS; S and HS; F and LH; S and LH.

A diskbb contribution in a class is due to the presence of either a geometrically thin Keplerian disc (Shakura \& Sunyaev 1973) or a highly radiation trapped geometrically slim disc (Abramowicz et al. 1988); the 
latter being
actually more suitably modeled by diskpbb. Being geometrically thin in nature with efficient well-understood cooling processes (thermal multicolor blackbody), a Keplerian disc, at moderately high accretion rates (but below the Eddington limit), is expected to reveal coherent flux in time which might render a deterministic and hence chaotic signature in its time series (or atleast a low $D_2$ or F). Note that chaos corresponds to the deterministic nature of the system and Keplerian discs also correspond to well determined cooling process balancing viscous heating, hence the link. This class of flow follows a single well-defined cooling mechanism in the optically thick regime above a certain value of mass accretion rate, making locally generated heat to be radiated out completely. 
Therefore, the appearance of a low $D_2$ in a pure DD (or HS) state argues for a Keplerian disc. However, the existence of a significant PL component does not allow the flow to be in a pure soft state. Appearance of a PL component must be due to the presence of an optically thin sub-Keplerian component additionally (see, e.g., Rajesh \& Mukhopadhyay 2010). Hence such flow may appear as the admixture of Keplerian disc (middle region) and sub-Keplerian halo (corona) (Chakrabarti \& Titarchuk 1995). Indeed, the classes $\beta,\lambda,\kappa$, while revealing low $D_2$, exhibit almost equal contributions from diskbb and PL, naming D-P states (see Table \ref{tab:tab1}). The class $\mu$ contains more diskbb, hence is close to the pure DD (or HS) state. However, these classes can also be understood as capturing the features of the repeated presence (high count-rate zone) and absence (dip) of the Keplerian disc with time, as easily understood from the corresponding lightcurves (Belloni et al. 1997, 2000). Hence, they reveal, as a whole (on average), moderately high fluxes. If the spectral state would be HS, then the flow 
plausibly is a pure Keplerian disc in a DD state with a very small $D_2$, the presence of a PL component increases $D_2$ (Mukhopadhyay 2004; Misra et al. 2004), leading it towards GAAF (see below). On the other hand, at a super-critical accretion rate, the flow becomes geometrically thicker and pressure supported (slim disc). With the increase of accretion rate, the flow thickness increases, making it quasispherical but radiation trapped, which 
may reveal incoherent flux in time (i.e. random behavior) in all directions (see, e.g., Abramowicz et al. 1988, Szuszkiewicz et al. 1996). This plausibly explains the appearance of Poisson noise like behavior (S) in the temporal classes along with a DD (or D-P) spectrum in certain flows ($\gamma$, $\phi$, $\delta$). 
In the present work, they in fact appear to be the admixture of slim disc and sub-Keplerian halo. 
In further analyses with more lightcurves, this heuristic argument needs to be confirmed. 
Nevertheless, using a diskbb model to fit a slim disc (which we do not know beforehand) appears
somewhat to be incompatible, as diskpbb model is more appropriate for slim discs. Hence, we have 
cross-checked our results for $\gamma$, $\phi$ and also $\delta$ classes with
diskpbb.
Interesting, using diskpbb model, disc contribution turns out to be even more than that of diskbb. 
For example, in the class $\gamma$, it increases to $79\%$ from $60\%$ and PL component decreases to 
$16.8\%$ from $31\%$. Also the Gaussian line contribution decreases to $4.2\%$ from $9\%$. More so, 
the inner disc temperature of this class increases slightly to 2.37keV (from 2.19keV) in line with 
the prediction of the diskpbb model. All of them argue more favorably the class to be a slim disc.
However, indeed the class $\gamma$ exhibits maximum flux, similar to, e.g. DD state of $\mu$. Nevertheless, due to the nature of highly trapping radiation presumably, fluxes of $\phi$, $\delta$ turn out
to be smaller. Moreover, the presence of significant PL component (sub-Keplerian halo) also destroys their nature of pure slim disc with higher fluxes. However, due to the increase of accretion rate, PL component of $\gamma$ is much smaller than that of $\beta, \lambda, \kappa$ and $\mu$ (Keplerian dominated flow) while $\delta$ seems to be in the transition phase between the Keplerian and slim discs. 
Nevertheless, by correlating time series and spectral analyses, $\gamma$ and $\phi$ classes appear more
like slim discs, which need to be analysed/re-confirmed in detail in future, may be with AstroSat data.

Out of five PD states, three ($\theta, \nu, \alpha$) are F. However, PD states correspond to effectively optically thin hot flows with hard photons, which must arise from a puffy sub-Keplerian flow. On the other hand, there is a class of sub-Keplerian optically thin flow, namely ADAF (Narayan \& Yi 1994), which exhibits a self-similar set of solutions, expected to be fractal in nature. Therefore, we argue that the classes with small $D_2$, revealing PD spectra, consist of ADAF. Such classes often show repeated outbursts/flares, captured in the lightcurves (see, e.g., Belloni et al. 2000) and hence reveal overall high count-rates and luminosities (except $\alpha$ which does not show outburst). ADAF is an extreme case of very low accreting systems with very insignificant cooling. As effectively there is no cooling, the ADAF and corresponding geometry is quite deterministic, based on a well-defined heating mechanism, with less uncertainly in the flow mechanism. This renders the flow to be hot, quasispherical, prone to produce outflows, independent of accretion rate below a certain critical value. This reveals the flux of the classes, with repeated ADAF and outburst, to be coherent in time and hence 
F in time series. The class $\alpha$ is almost pure ADAF with $77\%$ PL component and 
the lowest flux among all classes (indicative of a low accretion rate, as ADAF does). However, due to the presence of finite diskbb component, overall the class exhibits a nonnegligible flux. The time series of class $\chi$, however, reveals S along with a LH state. This nature emerges with the increase of accretion rate in ADAF --- compare the fluxes in $\alpha$ and $\chi$ classes. As mass accretion rate increases, density of the flow 
increases, which switches on certain cooling mechanisms randomly, leading to uncertainty and hence stochasticity in the flow physics. As the flow deviates from ADAF, it may start becoming less prone to produce outflows. The class $\chi$ also does not
exhibit any outburst. 
However, unless accretion rate and then cooling efficiencies are high enough, the flow remains puffed-up, quasi-spherical, optically thin, PD state, which is sub-Keplerian in nature (GAAF; Rajesh \& Mukhopadhyay 2010). As a result of random cooling processes, arising due to, e.g., Bremsstrahlung, synchrotron, inverse-Compton effects, the photon flux received by observers appears to be incoherent in time, leading the flow to be undeterministic or stochastic.

The class $\rho$ exhibits NS/LC (Mukhopadhyay 2004; Misra et al. 2004; Misra et al. 2006) 
with a very periodic X-ray lightcurve (see, e.g., Belloni et al. 2000). See, Misra et al. (2006) for detailed understanding of how this class is distinguished compared to other classes exhibiting F or S. On the other hand, it exhibits $72\%$ PL contribution revealing a PD state with a reasonably steep SI $\sim 2.9$. Hence it appears to be in between GAAF and Keplerian flow with a moderate luminosity (and accretion rate) in relative terms. The underlying flow might be low viscous (as generally might be the case for non-Keplerian flows) so that the infall time is very high. Hence infalling matter revolves locally in a longer period, which plausibly produces (almost) a periodic lightcurve. As the rate of supply of matter is steady, low viscous matter piles up locally and produces outflow.

All the above classifications can also be verified by the respective iron-line features shown by
the Gaussian line contribution. The iron line is expected to be caused by fluorescence
due to the irradiation of a Keplerian disc with hard X-rays from non-Keplerian part(s) (corona or/and 
sub-Keplerian zone). Hence, if
the Keplerian flow extends upto inner region exhibiting DD state, then the line should be broadened
due to general relativistic effects. This is seen in $\mu$ class as revealed by Fig. \ref{fig:fig1}
top-left panel. On the other hand, the ADAF class $\theta$ exhibits a smaller iron line contribution
showing a narrow line in Fig. \ref{fig:fig1} top-right panel, as the Keplerian region resides to be 
away from the black hole. Generally, with the increase of PL component, iron line features start becoming
insignificant. Nevertheless, for $\lambda$ and $\kappa$ classes (which somewhat exhibit 
intermediate spectral states), the best fit does not reveal any significant iron line feature. While
this behaviour is not very clear to us, which may be due to hindrance of irradiation of the Keplerian disc.
Note also that the moderate spectral resolution of RXTE might not allow us to make conclusive statements
regarding this behaviour, which needs to be verified further maybe by AstroSat data.
Nevertheless, the class $\chi$, although a PD class, still has an intermediate iron-line contribution with 
its intermediate
shape shown in Fig. \ref{fig:fig2} top-left panel, which confirms its GAAF nature. 
This is because of the presence of non-insignificant disc component leading to some fluorescence 
(although with less probability). The class $\delta$ also
seems to be an intermediate class, but close to slim disc, with 
quite a broadened iron-line, as shown in Fig. \ref{fig:fig2} top-right panel. 
This is because of its significant diskbb contribution from the vicinity of 
black hole.

\subsubsection{Model independent analysis}

\begin{figure*}
\begin{tabular}{@{}cc@{}}
\includegraphics[width=0.45\textwidth, height=0.33\textheight,angle=0]{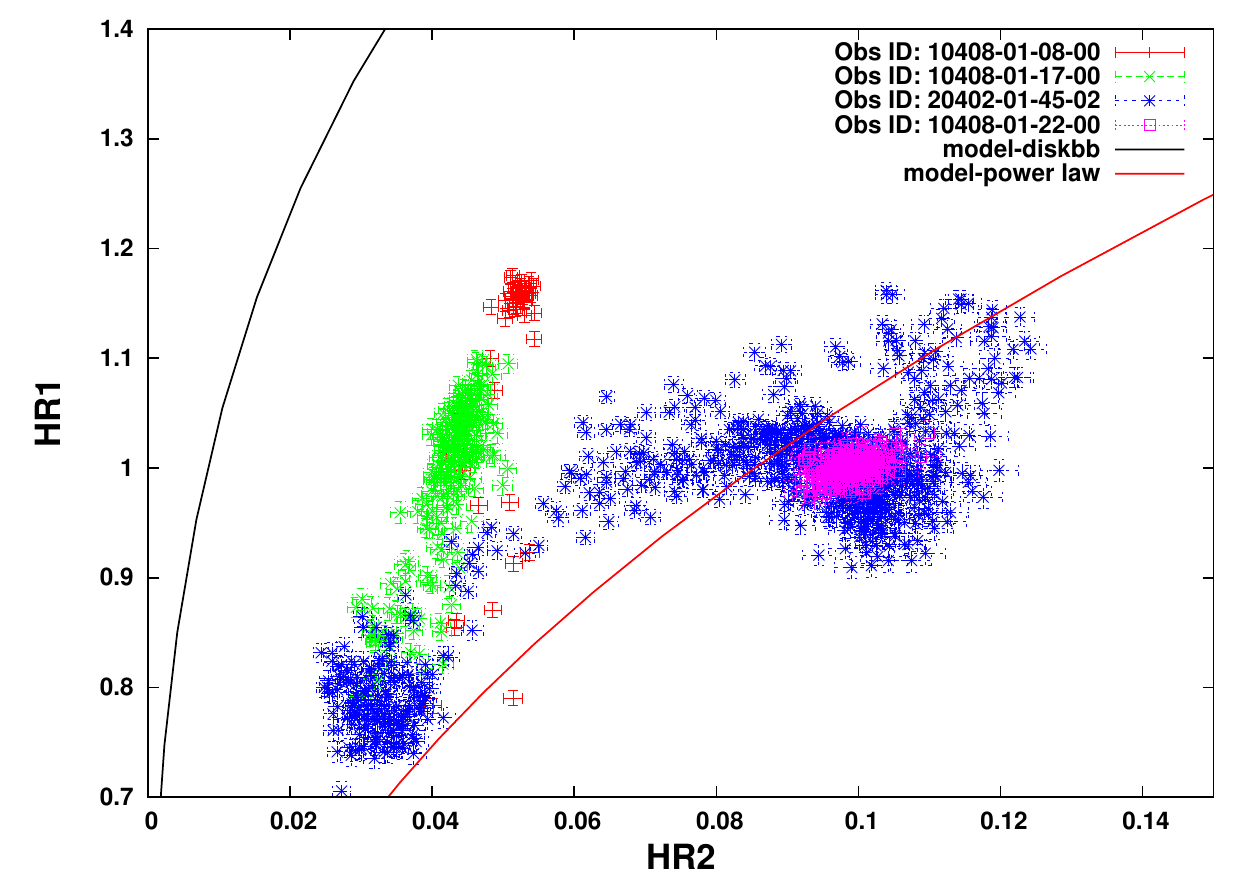} &
\includegraphics[width=0.45\textwidth, height=0.33\textheight,angle=0]{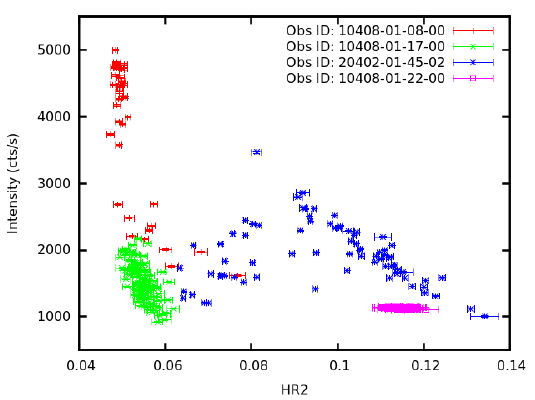}
\end{tabular}
\caption{The left panel represent the Color-Color diagram for the classes $\mu$ (solid `$+$' -red points), $\theta$ (short-dashed `$*$' -blue points), $\delta$ (dashed `$\times$' -green points) and $\chi$ (dotted `$\square$' -pink points), given in Table~1. The black and red solid lines represent the Color-Color diagram obtained from pure diskbb and powerlaw models respectively. See text for details. The right panel is the Hardness-Intensity diagram for the classes $\mu$ (`$+$' -red points), $\theta$ (`$*$' -blue points), $\delta$ (`$\times$' -green points) and $\chi$ (`$\square$' -pink points), given in Table~1. See text for details.
}
\label{fig:fig3}
\end{figure*}

It is important to double-check the spectral properties of our sources based on a model-independent approach. Therefore, in order to understand non-subjective properties, we
study the color-color diagram (CD) and hardness-intensity diagram (HI) of each class using \textit{RXTE} PCU standard $2$ data. Following previous work (Belloni et al. 2000), we define three energy bands, namely A ($2-5$ keV), B ($5-13$ keV) and C ($13-60$ keV). Then 
the hard color is defined as HR2=C/A and the soft color as HR1=B/A. In Fig. \ref{fig:fig3} we
show the CDs and corresponding HI diagram for four distinct temporal classes (one for each of the accretion modes considered in this work) described in Figs. \ref{fig:fig1}-\ref{fig:fig2}. 
It should be noted that the black (upper) and red (lower) lines respectively describe the CDs obtained from pure diskbb and PL models.
Note also that different points of the model diskbb curve correspond to different
inner disc temperatures and that of PL curve correspond to different powerlaw SIs. 

It is then verified that $\mu$ and $\chi$ classes, according to the present definition of color, indeed represent VH state with DD spectrum and LH state with PD spectrum respectively (with the former exhibiting the highest count-rates). However, the class $\delta$, owing to its smaller diskbb contribution and steeper SI compared to that in $\mu$, extends to the region of smaller HR1 with smaller count-rates. Finally, class $\theta$ extends from a region with large HR1 and HR2 to small HR1 and HR2 with intermediate count-rates. 
As the class exhibits a small diskbb contribution, it reveals a branch at small HR1 and HR2 (indeed, it is mostly clustered in the region of high HR1 and HR2). \\ \\

The above discussion in \S 3.1.4, along with that in \S 3.1.5, argues that GRS~1915+105 exhibits combinations of four classes of accretion flows over the time of evolution:
Keplerian disc (DD state with F), slim disc (DD state with S), ADAF (PD state with F) and GAAF (PD state with S). Although the lightcurves might not exhibit four pure classes, the combination of them may do. This plausibly confirms the existence of multicomponent accretion flows from observed data. This investigation is different than those done earlier (e.g. Belloni et al. 2000) which looked at the transitions within a lightcurve itself based on models (Keplerian and corona), hence they could not really comment on multicomponent nature of flows. The change of accretion rate switches the system from one class to another, which sometimes reveals intermediate (D-P) states, as discussed above based on the respective fluxes given in Table~1. If this is a generic picture, then other sources, e.g. neutron stars, also should exhibit similar/same trend correlation between nonlinear and spectral natures. 

\begin{figure*}
\centering
 \begin{tabular}{@{}cc@{}}
  \includegraphics[angle=270,width=.50\textwidth]{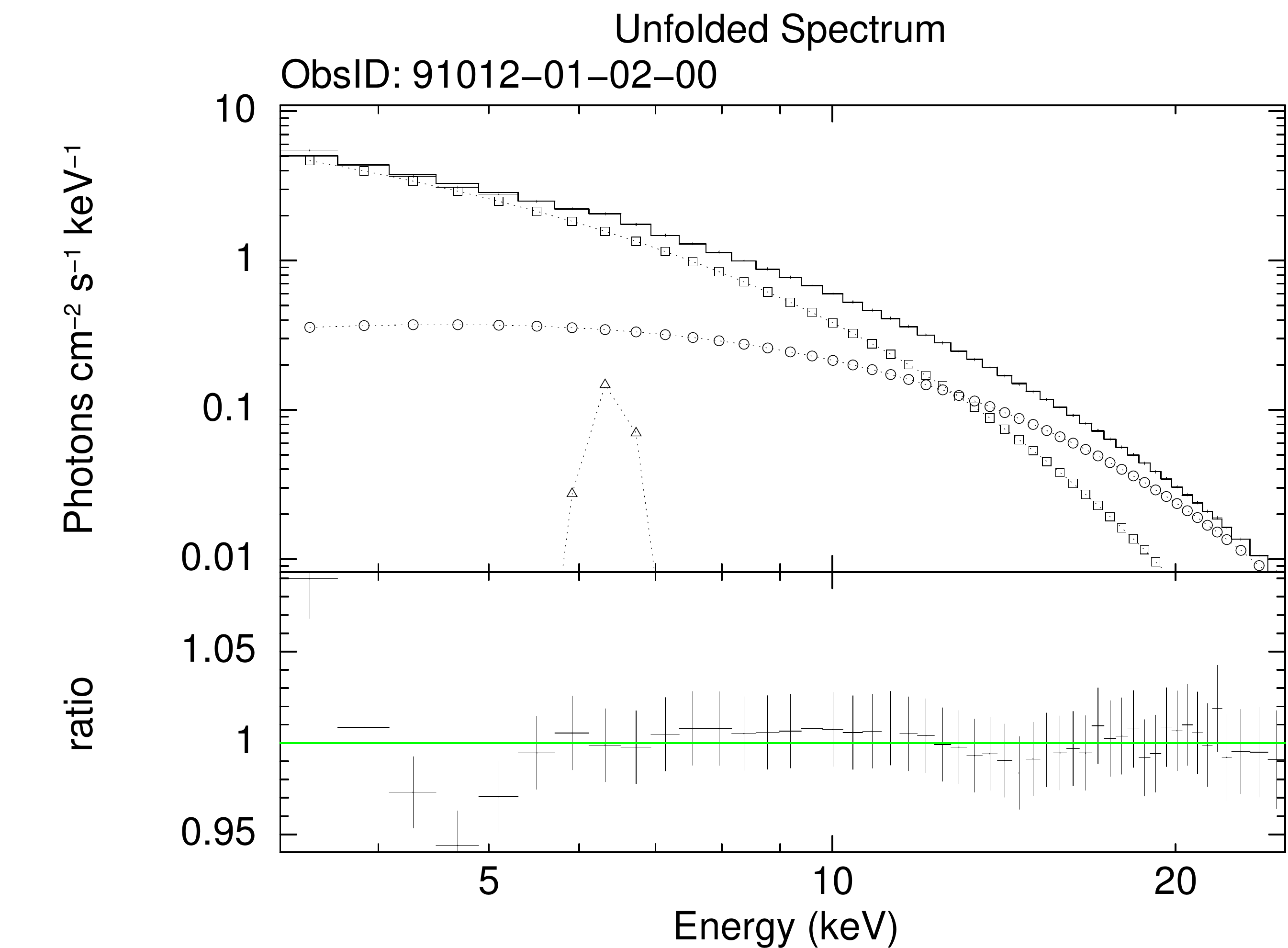} &
  \includegraphics[angle=270,width=.50\textwidth]{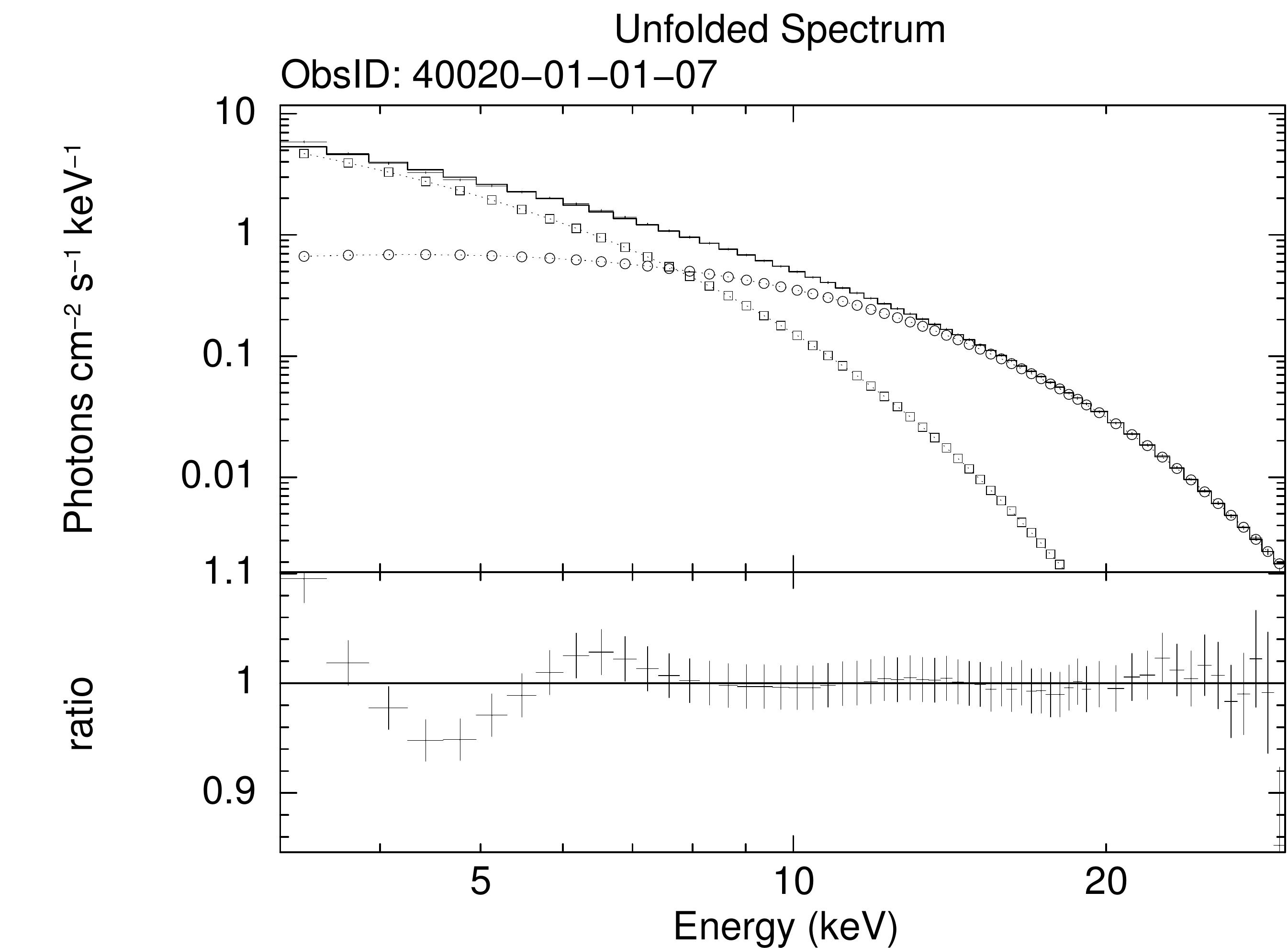} \\
  \\
  \includegraphics[width=.52\textwidth]{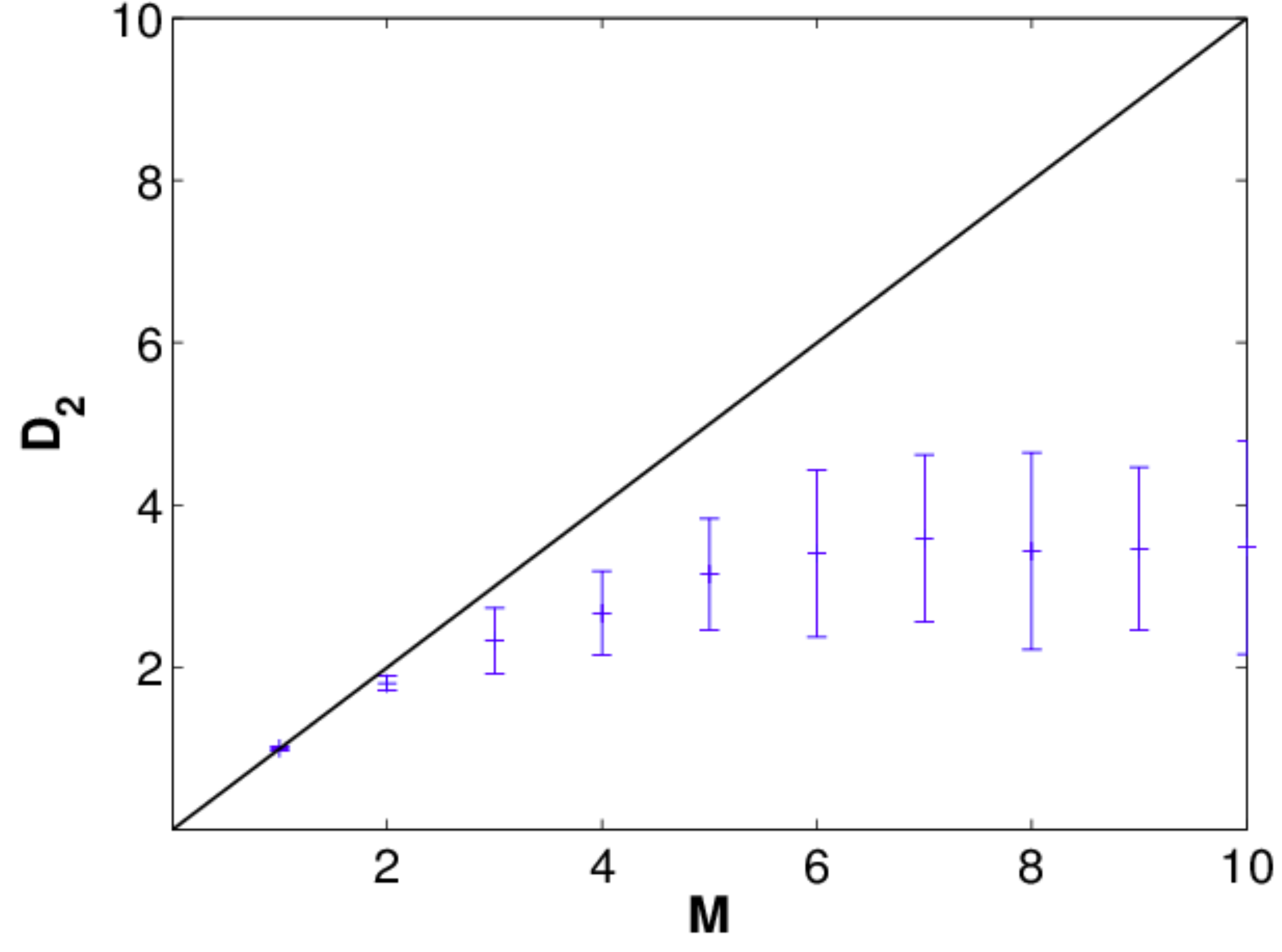} &
  \includegraphics[width=.52\textwidth]{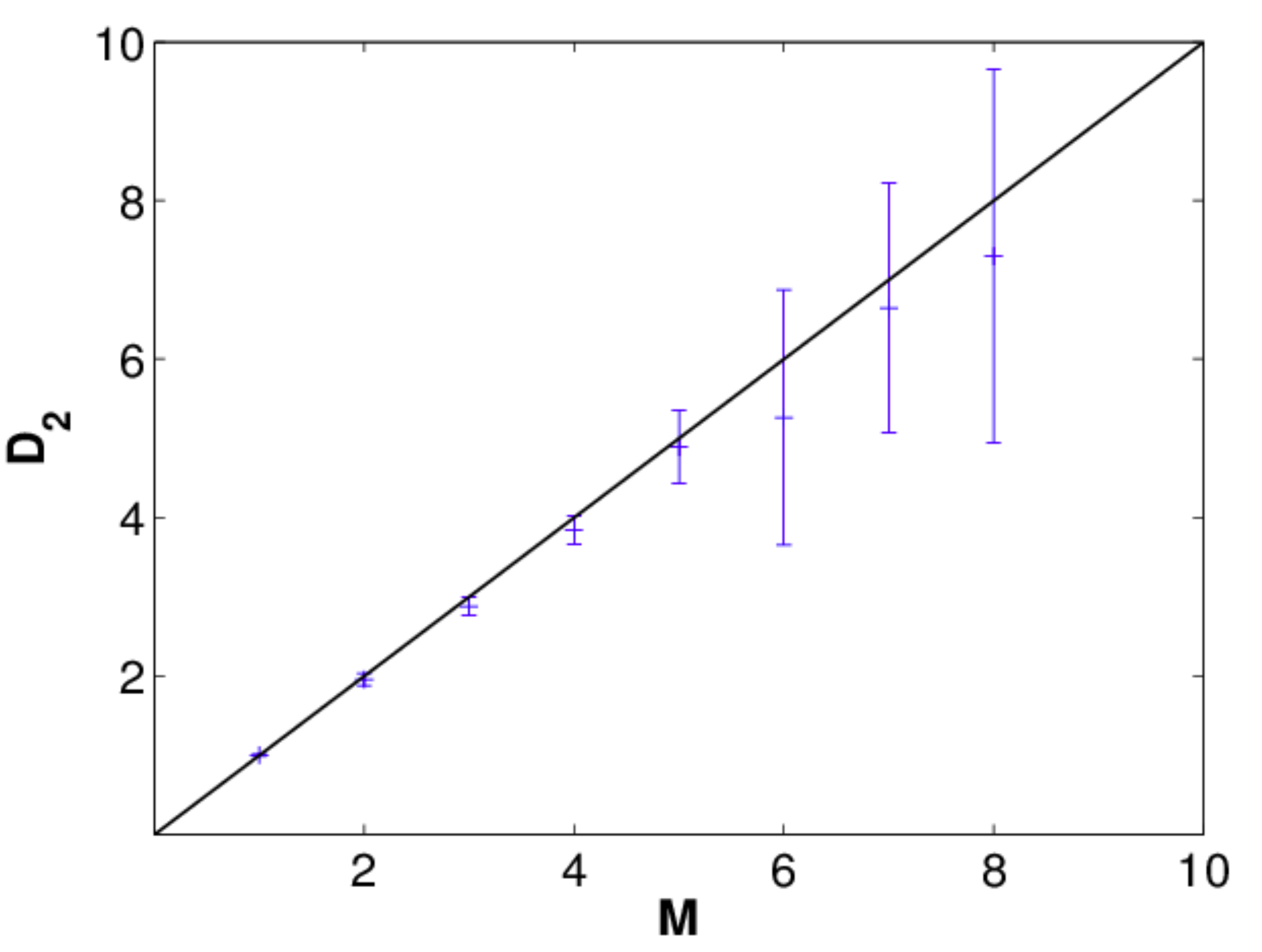} \\
 \end{tabular}
\caption{
Top panels: Unfolded spectra for ObsIDs 91012-01-02-00 (top-left) and 40020-01-01-07 (top-right) of Sco~X-1. The curves with square, circle and triangle symbols represent respectively diskbb, blackbody and Gaussian (iron) line contributions and the topmost solid curve in each panel represents the overall spectrum.
Bottom panels: Variation of correlation dimension as a function of embedding dimension with errors for ObsIDs 91012-01-02-00 (bottom-left) and 40020-01-01-07 (bottom-right) of Sco~X-1. The solid diagonal line in each panel represents the ideal stochastic curve. See Table~2 for other details.
}
\label{fig:fig4}
\end{figure*}

\begin{figure*}
\begin{tabular}{@{}cc@{}}
\includegraphics[width=0.45\textwidth, height=0.33\textheight,angle=0]{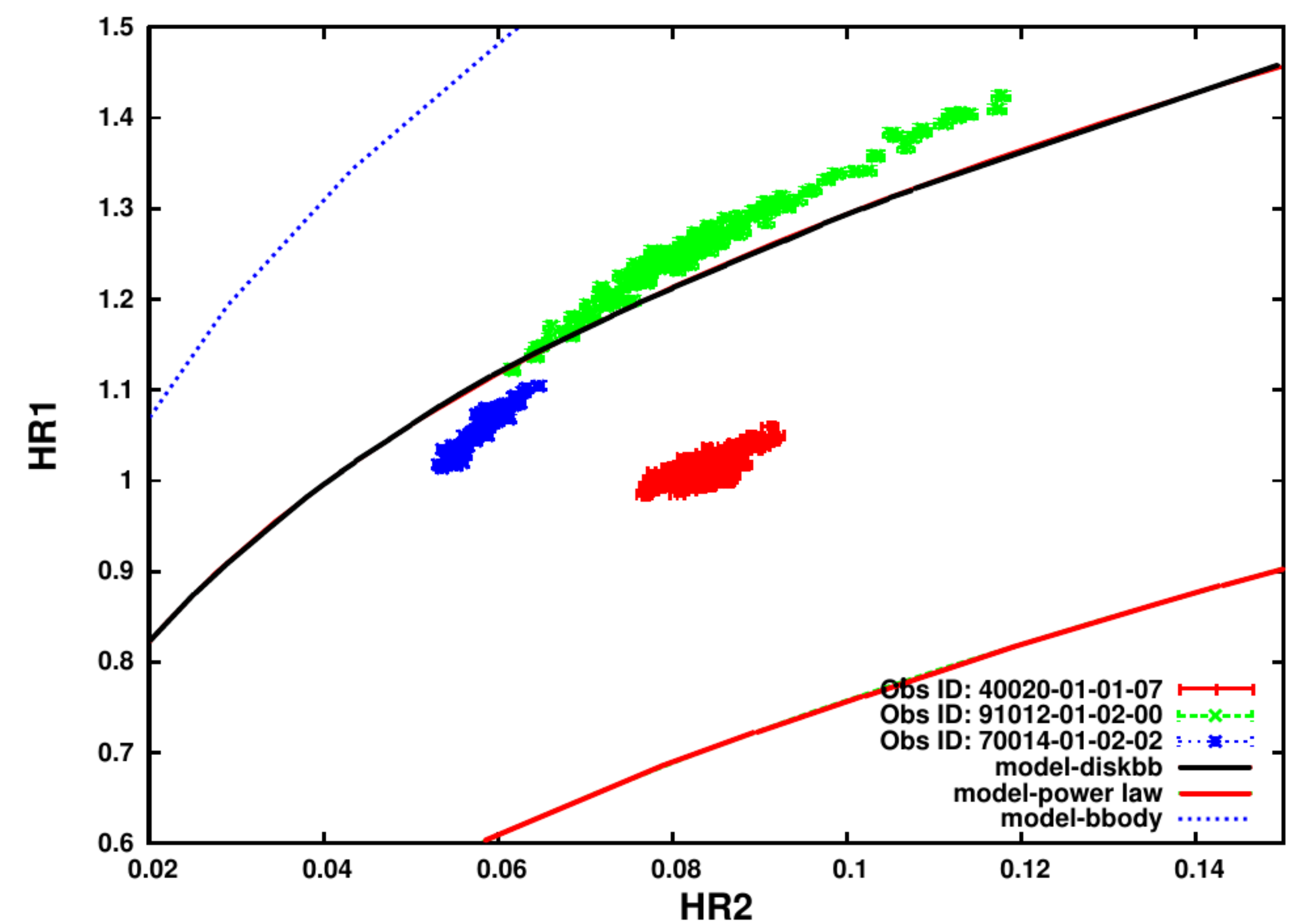} &
\includegraphics[width=0.45\textwidth, height=0.33\textheight,angle=0]{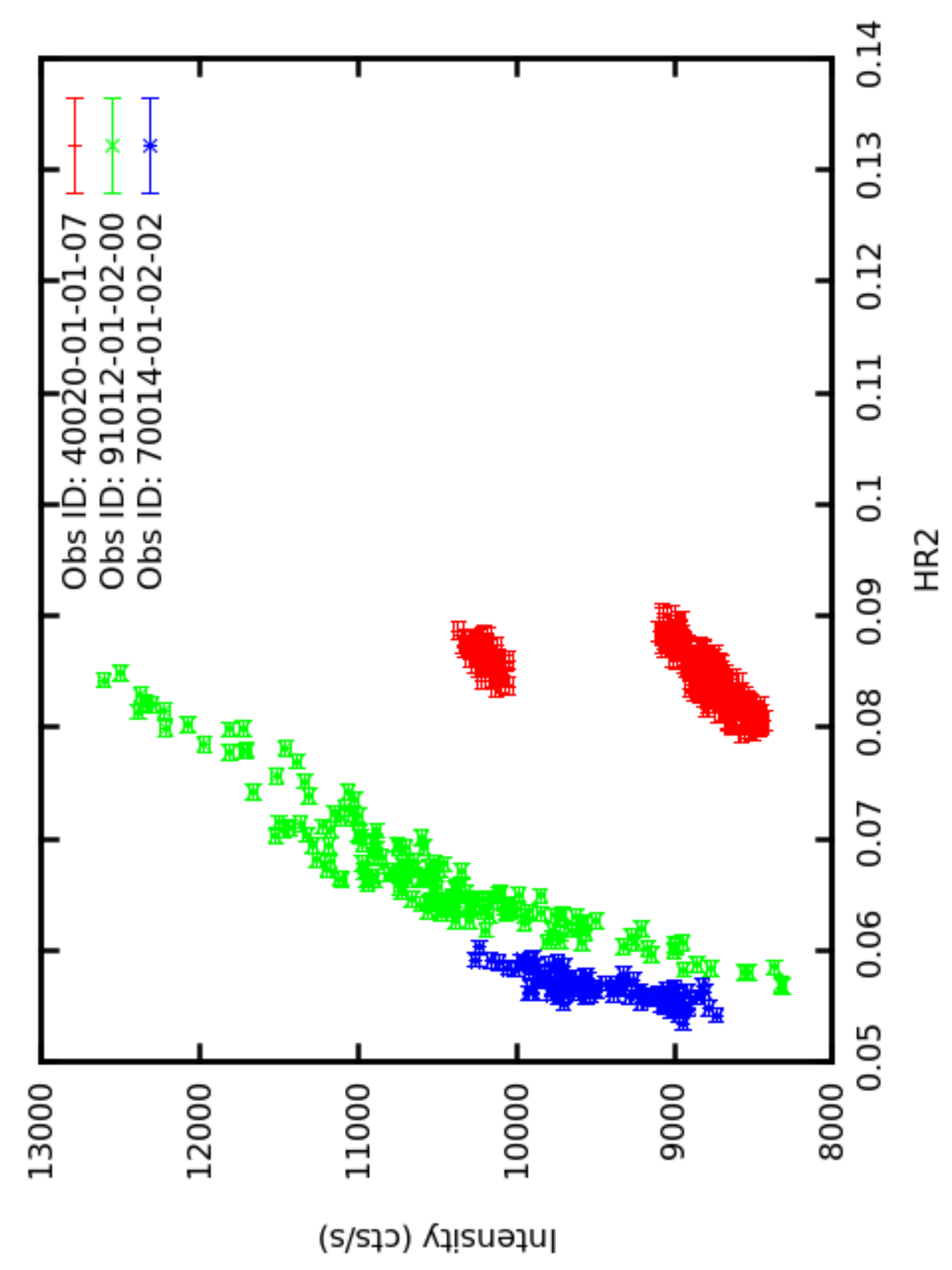} \\
\end{tabular}
\caption{The left panel shows the Color-Color diagram for the three different ObsIDs of Sco~X-1, given in Table~2. The upper, middle and lower diagram points (with `$\times$' -green, `$*$' -blue and `$+$' -red points respectively in the color version of the figure) correspond to ObsIDs 91012-01-02-00, 70014-01-02-02 and 40020-01-01-07 respectively. The upper, middle and lower (blue, black and red respectively in the color version of the figure) solid lines represent the Color-Color diagram plot obtained from pure blackbody, diskbb and power law (added for completeness) models. See text for details. The right panel is the Hardness-Intensity diagram for three different ObsIDs of Sco~X-1, given in Table~2. The diagram points with `$\times$' -green, `$*$' -blue and `$+$ -red points respectively correspond to ObsIDs 91012-01-02-00, 70014-01-02-02 and 40020-01-01-07.
}
\label{fig:fig5}
\end{figure*}


\subsection{Sco~X-1}

Karak, Dutta \& Mukhopadhyay (2010) already discussed the nonlinear properties of the time series of
neutron stars Sco~X-1 and Cyg~X-2 and found them to be chaos and/or of low $D_2$ sources. 
However, with further analysis, Sco~X-1 is found to deviate from its low $D_2$ nature reported earlier. 
Table \ref{tab:tab2} shows that the time series of Sco~X-1 have
temporal variabilities with three different classes. However, all of them seem to exhibit DD (and HS or VH) state. 
Note also that there is an additional blackbody (BB) component in the spectra. Indeed, the accretion flows between black hole and neutron star systems are different, primarily due to the presence of a surface in the latter (e.g., see Z-sources, Atoll sources 
and their associated color-color diagrams (Hasinger \& van der Klis 1989; Muno et al. 2002)). In addition the neutron star systems are known to generate less powerful radio outflows in comparison to the black hole systems (Kuulkers et al. 1997). 


Figure \ref{fig:fig4} describes the unfolded energy spectra, along with the corresponding nonlinear behavior of the time series, for ObsIDs 91012-01-02-00 and 40020-01-01-07, in order to understand various energy contributions to the spectra. Now correlating temporal/nonlinear and spectral behaviors, based on the discussion in \S3.1, we argue 
ObsID 91012-01-02-00 (first one in Table~2) for a Keplerian disc dominated flow and 40020-01-01-07 (last one in Table~2) for a slim disc dominated flow (with, however, lowest diskbb contribution making overall luminosity lowest). Like the slim disc classes of GRS~1915+105, here also we have cross-checked our results
with diskpbb, which further increases the disc contribution to $61.2\%$ from $55\%$ of diskbb fit 
and the BB component decreases to $38.8\%$ from $45\%$. Addition of an Gaussian (iron) line does not 
improve the fit and so it is not added. Also ObsID 70014-01-02-02 seems to be in the transition phase between them, when accretion rate starts increasing from a critical to super-critical values (however letting radiation to be trapped, like the slim disc cases of GRS~1915+105).
Note that BB component is expected to arise from the surface of the neutron star, which is absent for a black hole: GRS~1915+105. At a super-critical accretion rate, the thermal energy produced at the surface of a neutron star is high, leading to a very high contribution to BB in slim discs (ObsID 40020-01-01-07), reducing the relative diskbb contribution. 

This is also seen from CDs and HIs shown in Fig. \ref{fig:fig5}. The top, middle and bottom lines indicate CDs obtained respectively from pure BB, diskbb and PL models (for completeness). 
Interestingly, CD for 91012-01-02-00 lies very close to the model diskbb line exhibiting its Keplerian disc nature with high count-rates at low/intermediate energies as shown in HI. Its CD extends towards large HR2, indicating the presence of some hard photons as well in the source. Therefore, this ObsID might be in VH state, rather than HS state. CD and HI for 40020-01-01-07, which has a lower diskbb contribution than the former one, 
exhibit intermediate state between diskbb and PL obviously. A super-critical supply of matter here might contain hard photons and also exhibit high thermalization at the surface. Owing to its intermediate BB contributions, CD for 70014-01-02-02 lies between those of 91012-01-02-00 and 40020-01-01-07 and the corresponding HI appears to reveal a HS state.

\begin{table*}
\begin{center}
\caption{Sco~X-1 -- see Table \ref{tab:tab4} in Appendix for best fit parameters}
\label{tab:tab2}
\small
\begin{tabular}{cccccccccccccccccccccc}\\
\hline
\hline
 ObsID &  behavior & diskbb & BB & GA & $\chi^{2}/\nu$ & state & count-rate & $L$ \\
\hline
\hline
91012-01-02-00  & F & $74$ & $25$ & $01$ & $0.8202$ & HS/DD/VH & 15827 & $0.9249$\\
 70014-01-02-02  & NS & $65$ & $39$ & $01$ & $0.7344$ & HS & 15011 & $0.8462$\\
 40020-01-01-07  & S & $55$ & $45$ & $-$ & $0.9701$ & HS & 14554 & $0.8422$\\
\hline
\hline

\end{tabular}
\end{center}
{Columns:- 
1: RXTE observational identification number.
2: The behavior of the system (F: low correlation/fractal dimension;
S: Poisson noise like stochastic; NS: non-Poissonian).
3: $\%$ of multi-color disc blackbody component.
4: $\%$ of blackbody component.
5: $\%$ of Gaussian line component (XSPEC model gauss).
6: Reduced $\chi^2$.
7: Spectral state (HS: high/soft; VH: very high; DD: disc dominated).
8: Model predicted count-rate (cts/s).
9: Total luminosity in $3-25$ keV in units of Eddington luminosity for the neutron star mass $1.4M_\odot$
(Bradshaw, Fomalont \& Geldzahler 1999; Steeghs \& Casares 2002).
}
\end{table*}

\subsection{Results from simulated lightcurve}
In order to strengthen our inference discussed above, we explore simulated lightcurves and corresponding time series. In this connection, we concentrate on a particular class of flow, ADAF, and perform all the analyses on it as done with observed lightcurves. The method and result are described as follows

\subsubsection{Equations to be solved}
The numerical set-up described here is used to obtain the lightcurve from an advection dominated accretion flow around a black hole. 
In our numerical simulations, we solve Newtonian magnetohydrodynamic (MHD) equations in spherical co-ordinates ($r,~\theta,~\phi$) 
using {\tt PLUTO} code (Mignone et al. 2007). The equations are
 \begin{eqnarray} 
 \label{eq:mass}
&& \frac{\partial \rho}{\partial t} + \nabla .(\rho \textbf{v})= 0, \\
\label{eq:momentum}
&& \frac{\partial }{\partial t}\left(\rho \textbf{v}\right)+ 
\nabla . \left( \rho \textbf{v}\textbf{v} - \textbf{B}\textbf{B} \right)=  -\rho \nabla \Phi - \nabla P^*, \\
\nonumber
&& \frac{\partial}{\partial t}(E +\rho\Phi) + \nabla . \left[(E + P^* +\rho\Phi)\textbf{v} - \textbf{B}(\textbf{B}.\textbf{v})   \right]= 0, \\ 
\label{eq:energy}
\\
\label{eq:induction}
&& \frac{\partial \textbf{B}}{\partial t} + \nabla . \left (\textbf{v} \textbf{B} - \textbf{B}\textbf{v} \right) = 0,
 \end{eqnarray}
where $\rho$ is the gas density, $\textbf{v}$ is the velocity vector, $\textbf{B}$ is the magnetic field vector (a factor of $1/\sqrt{4 \pi}$ is absorbed in the definition of $\textbf{B}$), and $E$ is the total energy density. 
$P^*$ is the total pressure given by, $P^* = P + B^2/2$; where, $P$ is  gas pressure. $E$ the total energy density is related to internal energy 
density $\epsilon$ as $E = \epsilon + \rho v^2/2 + B^2/2$. To mimic the  general relativistic effects close to the black hole,
we use pseudo-Newtonian potential proposed by Paczynski \& Wiita (1980), $\Phi = GM/(r-2r_g)$, where $r_g = GM/c^2$ is the gravitational radius,
$M$, $G$ and $c$ are the mass of the accreting black hole, gravitational constant and speed of light in vacuum respectively. We work in dimensionless units, where $GM=c=1$. 
Therefore,  in this section all the length and time scales are given  in units of $GM/c^2$ and $GM/c^3$ respectively, unless stated otherwise. 

{\tt PLUTO} uses a Godunov type scheme  which follows a conservative ansatz. Flock et al. (2010) extensively used {\tt PLUTO}
to study global simulations of magnetorotational instability (MRI) and showed that a particular combination of the choices of algorithmic is
necessary to properly capture the linear growth of MRI. Following the recommendation of  Flock et al. (2010), we use the HLLD (Miyoshi \& Kusano 2005)
solver with second-order slope limited reconstruction. For time-integration, second order Runge-Kutta (RK2) is used with CFL number 0.3. 
For the calculation of the electromotive forces (EMFs) for the induction equation, we use the upwind-constrained transport `contact' method by 
Gardiner \& Stone (2005).
\subsubsection{Initial conditions}
We initialize with an equilibrium solution given by Papaloizou \& Pringle (1984), which describes a constant angular momentum torus embedded 
in a non-rotating, low density medium  in hydrostatic equilibrium. Pressure and density within the torus follow a polytropic equation of state $P = K \rho^{\gamma}$, $K$ is a constant. Following Das \& Sharma (2013), we determine the initial density profile of the torus given by,
\begin{eqnarray}
\label{torus_den}
\rho^{\gamma -1} = \frac{1}{(n+1)R_0 K} \left[ \frac{R_0}{r-2} - \frac{1}{2}\frac{R^4 _0}{[(R_0-2)R]^2} - \frac{1}{2d} \right ]
\end{eqnarray}
where $\gamma=5/3$ is the adiabatic index, $n = 1/(\gamma -1)$ is the polytropic index, $R=r {\rm sin} \theta$ is the cylindrical radius and $R_0$ 
is the cylindrical radial distance of the center of the torus from the black hole. $d$ is the distortion parameter which determines the shape and size of the torus. 
The density of the torus is maximum ($\rho = \rho_0$) at $r=R_0$ and $\theta=\pi/2$. This information is used to calculate $K$ and given by
\begin{eqnarray}
\label{ken}
K = \frac{1}{(n+1)R_0 \rho^{\gamma-1}_0} \left[ \frac{R_0}{R_0-2} - \frac{1}{2}\frac{R^2 _0}{(R_0-2)^2} - \frac{1}{2d} \right].
\end{eqnarray}
The constant angular momentum $l_0$ associated with the torus is given by its Keplerian value at $R_0$, 
\begin{eqnarray}
\label{lkep}
l_0 = \frac{R^{\frac{3}{2}}_{0}}{R_0 - 2}.
\end{eqnarray}
We choose $\rho_0 = 10^6~m_p cm^{-3}$ where $m_p$ being the mass of proton, 
$R_0 = 20$ and $d=1.15$. The chosen value of $d$ makes the initial torus geometrically thick ($H/r \sim 0.1$ at $R_0$). We choose 
the density of the ambient medium to be small enough ($\rho_{\rm amb}=10^{-4}\rho_0$), such that it does not affect our results. It is to be mentioned that  
the choice of $\rho_0=10^6~m_p cm^{-3}$ is quite low for an advective flow in a black hole binary system. But the absolute value of $\rho_0$ does not really  
affect the dynamics of the flow, it is scaled out from the fluid equations as long as cooling term is not incorporated in  the energy equation.

We initialize a poloidal magnetic field which threads the initial torus and is parallel to the density contours. This magnetic field is defined through a vector potential
\begin{eqnarray}
\label{vec}
A_{\phi} = C\rho^2,
\end{eqnarray}
which guarantees the divergence free nature of the magnetic field. $C$ is a constant which determines the field strength. The initial magnetic field strength is quantified by the ratio of average (over volume) gas ($P_V$) to magnetic ($B^2_V/8\pi$) pressures 
and is defined by
\begin{eqnarray}
\label{beta}
\beta_{\rm ini} = \frac{P_V}{B_V^2/2},
\end{eqnarray}
where the volume average is done only over the initial torus.
We choose $\beta_{\rm ini} = 715$ in our simulation. 
\\
\subsubsection{Numerical set-up}
We carry out  three-dimensional (3D) MHD simulations in spherical co-ordinates ($r,~\theta,~\phi$). The computation domain extends from an 
inner radius $r_{\rm in}=4$ to an outer radius $r_{\rm out}=140$. We use a logarithmic grid along the radial direction from $r_{\rm in}$ to $r_{l}=45$
with $N_{rl}=536$ grid points. In the outer region, from $r_l=45$ to 
$r_{\rm out}$, we use a stretched grid with $N_{rs}=32$ grid points. This outer region acts as a buffer zone. Along meridional direction, a uniform grid with
$N_{\theta u} = 232$ is used for  $\pi/3  \leq \theta \leq  2\pi/3$. For $0 \leq \theta < \pi/3$ and 
$2\pi/3 \leq \theta \leq \pi$, we use stretched grid with $ N_{\theta s}=24$ grid points for each interval. In the azimuthal direction, 
to save computation cost, our model is constrained to $0 \leq \phi \leq \pi/4$ with $N_{\phi}=128$ grid points.

We use a pure inflow ($v_r\leq 0$) boundary condition at the radial inner boundary, so that nothing comes out off the black hole.
Zero gradient boundary conditions are used for $\rho$, $p$, $v_{\theta}$ and $v_{\phi}$.
At the radial outer boundary, we fix the density and pressure to their initial values, $v_{\theta}$ and $v_{\phi}$ are kept free,
but we set $v_r \geq 0$ so that matter can go out of the computation domain but can not come in. At both the  inner and outer radial boundaries, 
 `force-free' (zero-gradient) conditions are applied  for tangential components ($B_r$ and $B_{\theta}$) of $\textbf{B}$-field, while the normal 
 component of $\textbf{B}$-field  is determined by $\nabla . \textbf{B}=0$ condition. Reflective boundary conditions are applied at both 
 $\theta$-boundaries. We use periodic boundary conditions at both the azimuthal boundaries.

 \subsubsection{Results}

 \begin{figure}
  \centering
    \includegraphics[scale=0.35]{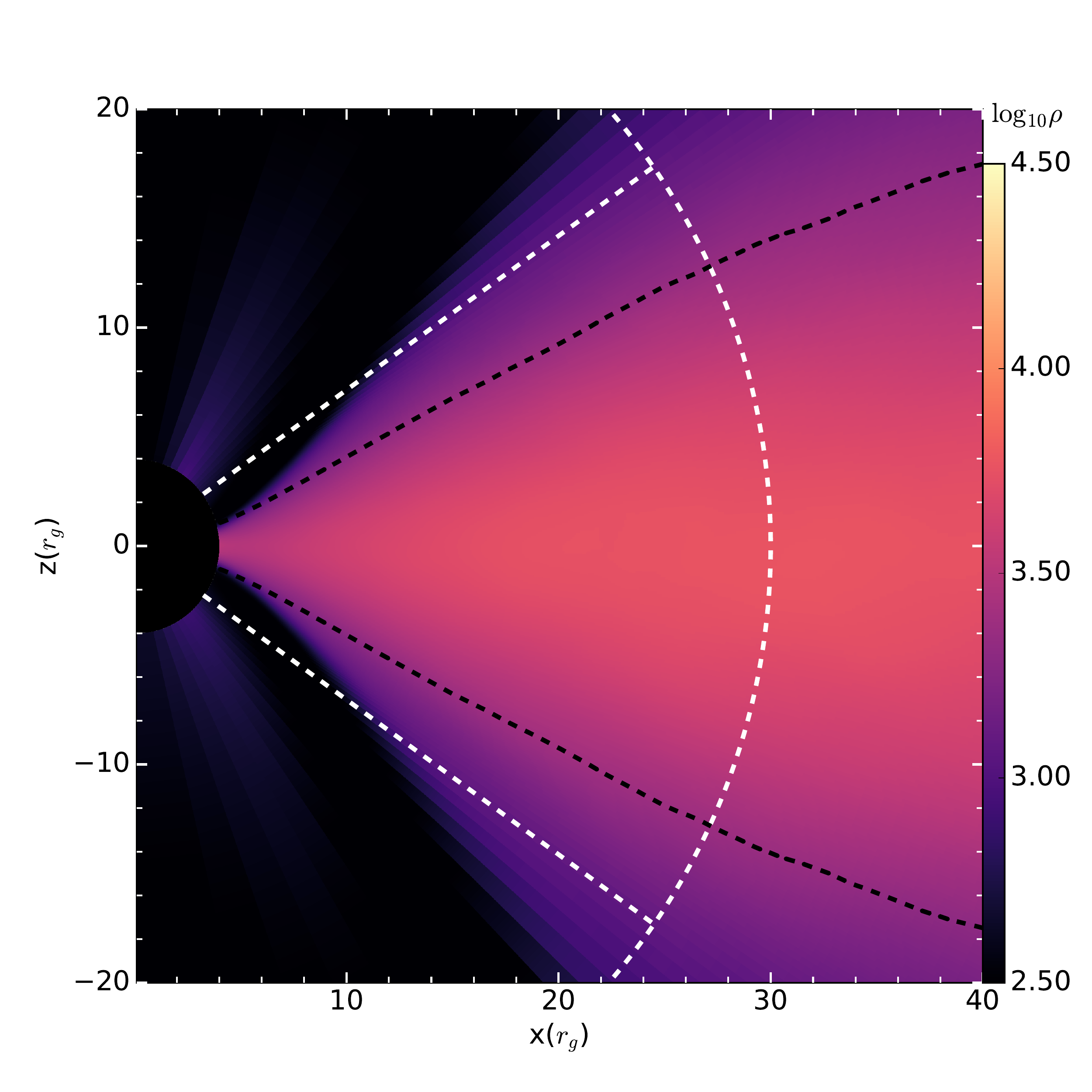}
     \caption{The average density profile $\bar{\rho}(r,\theta)$; average  is done over the azimuthal angle $\phi$  and over the time duration, we consider for 
     calculating lightcurve. Black dashed line describes the variation of scale  height $H$ with radial distance $r$.  Luminosity is calculated by taking the contribution from 
     the region confined within white dashed lines. }
    \label{fig:den_avg}
 \end{figure}
Figure \ref{fig:den_avg} shows the $\phi$-averaged and time-averaged density profile. Time average is done over the interval which we consider  for time-series analysis 
of lightcurve, i.e from $t_1=26000$ to $t_2=48000$. The black dashed line describes the scale height ($H$) above and below the 
mid plane ($\theta=\pi/2$); where scale-height $H$ is given by
\begin{eqnarray}
H = \frac{\bar{c}_s(r,\theta=\frac{\pi}{2})}{\bar{\Omega} (r,\theta=\frac{\pi}{2})},
\end{eqnarray}
where bar ($-$) represents the $\phi$-average and $c_s=\sqrt{\gamma P/\rho}$ and $\Omega$ are sound speed  and angular velocity  respectively.
Over a broad range of  radial distance $r$, $H/r$ is almost constant $\sim 0.4$, which implies that the accretion flow is geometrically thick.
For calculating the lightcurve, we choose the region,
enclosed by the white dashed line, in the radial direction which is extended from the inner boundary to a  radial distance  up to $r_0 = 30$ and in the meridional direction it is extended 
from $\theta=45^{\circ}$ to $\theta=135^{\circ}$. The region is chosen in such a way that both dense accretion flow near the mid plane  and  tenuous corona off the mid plane  contribute
to the lightcurve.

\subsubsection{Resolving MRI and inflow equilibrium}
Standard test of convergence is beyond the scope of this paper which is done on running simulations with increasing resolution while keeping the underlying problem unchanged. To check whether MRI is well resolved or not in our global simulation of accreting torus, we calculate three convergence metrics (Noble, Krolik \& Hawley 2010; Hawley, Guan \& Krolik 2011; Sorathia et al. 2012),
\begin{eqnarray}
\label{eq:metrics}
&& Q_{\theta}  = \frac{2\pi}{v_{\phi} \Delta \theta} \frac{|B_{\theta}|}{\sqrt{\rho}}, \\
&& Q_{\phi}  = \frac{2\pi}{v_{\phi} {\rm sin}\theta \Delta \phi} \frac{|B_{\phi}|}{\sqrt{\rho}},\\
&& \theta_B = \frac{1}{2}{\rm sin}^{-1} \left(\frac{-2B_r B_{\phi}}{B^2} \right),
\end{eqnarray}
where  quality factor $Q_{\theta}$ measures the number of cells in the $\theta$-direction across a wavelength of fastest growing mode,
$\lambda^{\rm MRI}_{\theta} = 2 \pi |B_{\theta}|/\Omega \sqrt{\rho} $ and $Q_{\phi}$ measures number of cells in the $\phi$-direction across a 
wavelength of fastest growing mode, $\lambda^{\rm MRI}_{\phi} = 2 \pi |B_{\phi}|/\Omega \sqrt{\rho}$.  $\Delta \theta$ and $\Delta \phi$ are the grid cell sizes in $\theta$- and $\phi$-directions respectively, $B_{\theta}$ and $B_{\phi}$ are the magnetic field components, $\Omega=v_{\phi}/r$ is the fluid's angular velocity. $\theta_B$ is the ratio of Maxwell stress ($-B_r B_{\phi}$) to  magnetic pressure ($B^2/2$) and defines a characteristic tilt angle of the magnetic field with respect to the toroidal direction (Guan et al. 2009; Pessah 2010). The magnetic tilt angle is related to the efficiency of angular momentum transport at a given field strength, and can be used to probe whether a simulation has correctly captured the nonlinear saturation of the MHD turbulence.

\begin{figure}
  \centering
    \includegraphics[scale=0.45]{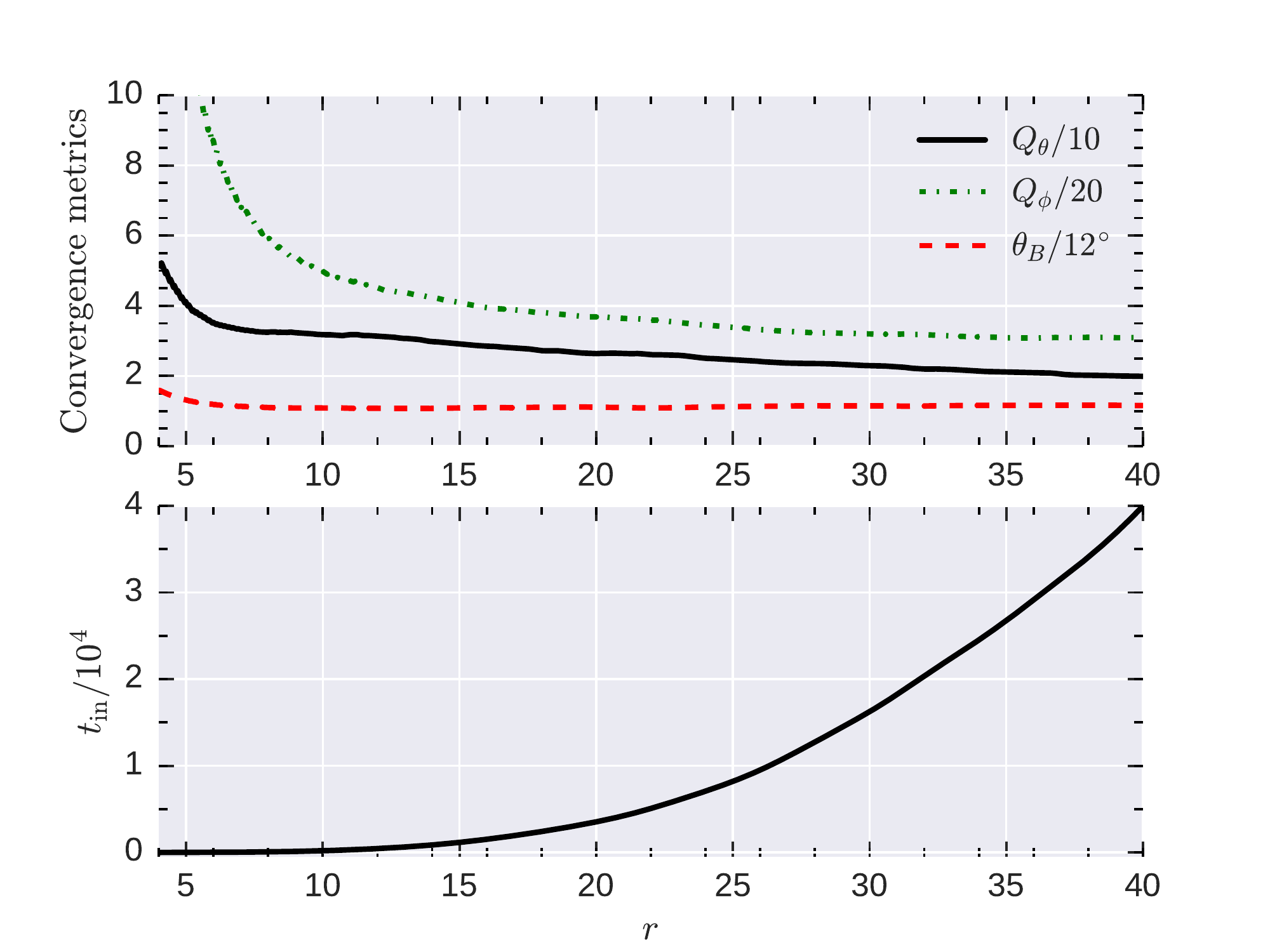}
     \caption{Top panel: Variation of different convergence metrics ($Q_{\theta}$, $Q_{\phi}$ and $\theta_B$) with radial distance $r$ from central accretor. The metrics 
     are spatially averaged over $\theta$, $\phi$ and time averaged from $t_1$ to $t_2$. Bottom panel: Radial infall time $t_{\rm in } = \int dr/v_r(r)$ at different $r$.}
    \label{fig:1d_prof}
 \end{figure}

According to Hawley, Guan \& Krolik (2011), simulations with $Q_{\theta} \geq Q_{\theta, c}=10$ and $Q_{\phi} \geq Q_{\phi, c}= 20$ are well resolved, and essential for simulating non-linear behavior. Also azimuthal resolution is coupled to meridional resolution, smaller $Q_{\phi}$ can be compensated for by larger $Q_{\theta}$. While quality factors 
give the information of how well MRI is resolved, they do not address the structure of turbulence in saturated state. Magnetic tilt angle $\theta_B$
above a critical value confirms the transition from linear growth of MRI to saturated turbulence and the critical value is $\theta_{B,c} \sim 12^{\circ}$ (Pessah 2010). 
The top panel of Fig. \ref{fig:1d_prof} shows the variation of  average convergence metrics (namely, $Q_{\theta}$, $Q_{\phi}$ and $\theta_B$)
with radial distance $r$. We normalize the average value of  $Q_{\theta}$, $Q_{\phi}$ and $\theta_B$ by the critical values ($Q_{\theta,c}$, $Q_{\phi,c}$ and 
$\theta_{B,c}$ respectively)
required for convergence. We take both  spatial (over $\theta$ and $\phi$) and temporal (over the interval from $t=26000$ to $t=48000$) average
of the metrics. In the meridional direction, we consider the contribution only from the region within one scale height (from  $+H$ to $-H$), while 
in the azimuthal direction, average is done over the entire available domain ($\phi=0$ to $\phi=\pi/4$).  In our simulation, we find that 
both $Q_{\theta}$ and $Q_{\phi}$ are much greater than their critical values, while the magnetic tilt angle is always close to $\theta_{B,c}$.
Hence, we conclude that in our simulation, MRI is well-resolved throughout 
the region of interest and has correctly captured the nonlinear saturation of the MHD turbulence. In addition, in our simulation, the ratio $r \Delta \phi/\Delta r \sim 1.35$ near the mid plane, which is sufficient [$(r \Delta \phi/\Delta r)_c\leq2$]
for convergence according to Sorathia et al. 2012.

We want to analyze the time series of lightcurve $L(t)$ (see Eq. (\ref{eq:lum}) below) obtained in a statistically stationary state. Now our global disc simulation starts with 
a finite mass in the computation domain; due to accretion most of it is transferred toward central accretor, while a small fraction of mass
moves outward carrying a large amount of angular momentum. As a result, only a part of the flow can be in `inflow equilibrium'  which has
had sufficient time to settle down to a statistically stationary state. There are several procedures for determining  `inflow equilibrium'. One way is to compare the simulation time with average infall time of a fluid element at radius $r$ (Hawley, Guan \& Krolik 2011). We compute time-averaged (between $t_1=26000$ to $t_2=48000$) radial velocity  $v_r(r)$ of the gas within one scale-height of the mid-plane. Using this, we calculate average
infall time as,
\begin{eqnarray}
t_{\rm in} = \int_{r_{\rm in}}^{r} \frac{dr}{v_r(r)}
\end{eqnarray}

The bottom panel of Fig. \ref{fig:1d_prof} shows the variation of infall time $t_{\rm in}$ as a function of $r$. The average infall time from
$r_0=30$ is $t_{\rm in}(30)\sim 1.6 \times 10^4$, which states that the region within $r_0=30$ attains `inflow equilibrium' by the time  we 
calculate the lightcurve. For time series analysis, we use the lightcurve $L(t)$ between time $t_1=2.6 \times 10^4$ to $t_2=4.8 \times 10^4$, 
which means the lightcurve is obtained when the accretion flow is in a statistically steady state.  

 \subsubsection{lightcurve}
In our simulation set-up  for a radiatively inefficient accretion flow, we do not incorporate any information about cooling.
Because, due to the fact that cooling time scale is much longer than the viscous time scale in radiatively inefficient accretion 
flow, cooling does not really affect the dynamics of the flow. To obtain the required lightcurve, we calculate however the free-free 
luminosity,
\begin{eqnarray}
\label{eq:lum}
L(t) = \int_{r_1}^{r_2} \int_{\theta_1}^{\theta_2} \int_{0}^{\pi/4} n^2 \Lambda r^2 {\rm sin}\theta dr d\theta d \phi
\end{eqnarray}
originated from the accretion flow within a specified region shown  in Fig. \ref{fig:den_avg} (region enclosed by white dashed line) in post-processing. Here, $n$ is the number density, $\Lambda \propto T^{1/2}$ is the cooling function, $T$ is the electron temperature of the flow. It is to be noted that we do not consider the (more realistic) contributions
due to inverse-Compton effect and synchrotron emission, which are difficult to model. But both the  cooling terms are expected to be $\propto n^2$,
as radiation and magnetic energy densities are likely to have a scaling with mass accretion rate. Therefore, Bremsstrahlung  cooling should give qualitatively 
similar results which is expected with the inclusion of inverse-Compton effect and synchrotron emissions. 
Also we use a single fluid model, whereas, the electrons in hot accretion flows are at
lower temperature compared to that of the protons. Therefore lightcurve from simulation should 
be only taken as trends, which however suffices for the present purpose.
We dump the data  frequently enough such that we have sufficient number of data points 
for time series analyses.

\begin{figure}
  \centering
    \includegraphics[scale=0.45]{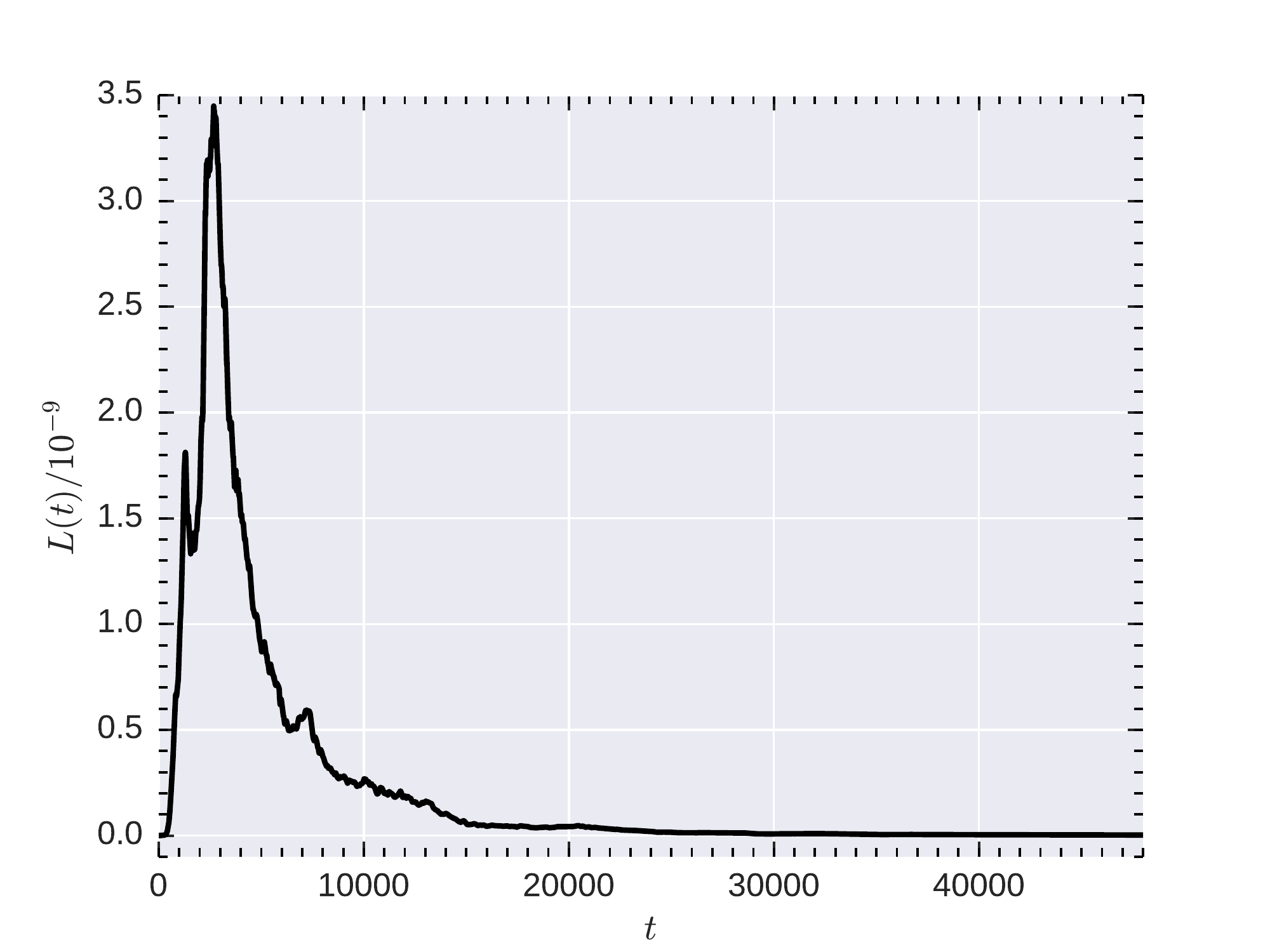}
     \caption{lightcurve $L(t)$ for whole duration of simulation run time. 
}
    \label{fig:lumin}
 \end{figure}

Figure. \ref{fig:lumin} shows the lightcurve $L(t)$ for the whole duration of simulation time. During the initial transient phase of accretion,
when initial torus expands due to magnetic shear and MRI sets in, which enables accretion of mass, luminosity coming out 
from the specified region reaches maximum. As the initial transient phase is over, flow attains a quasi-stationary state,  and $L(t)$
becomes quasi-steady and we consider that region for our analysis.

\begin{figure}
\includegraphics[scale=0.65]{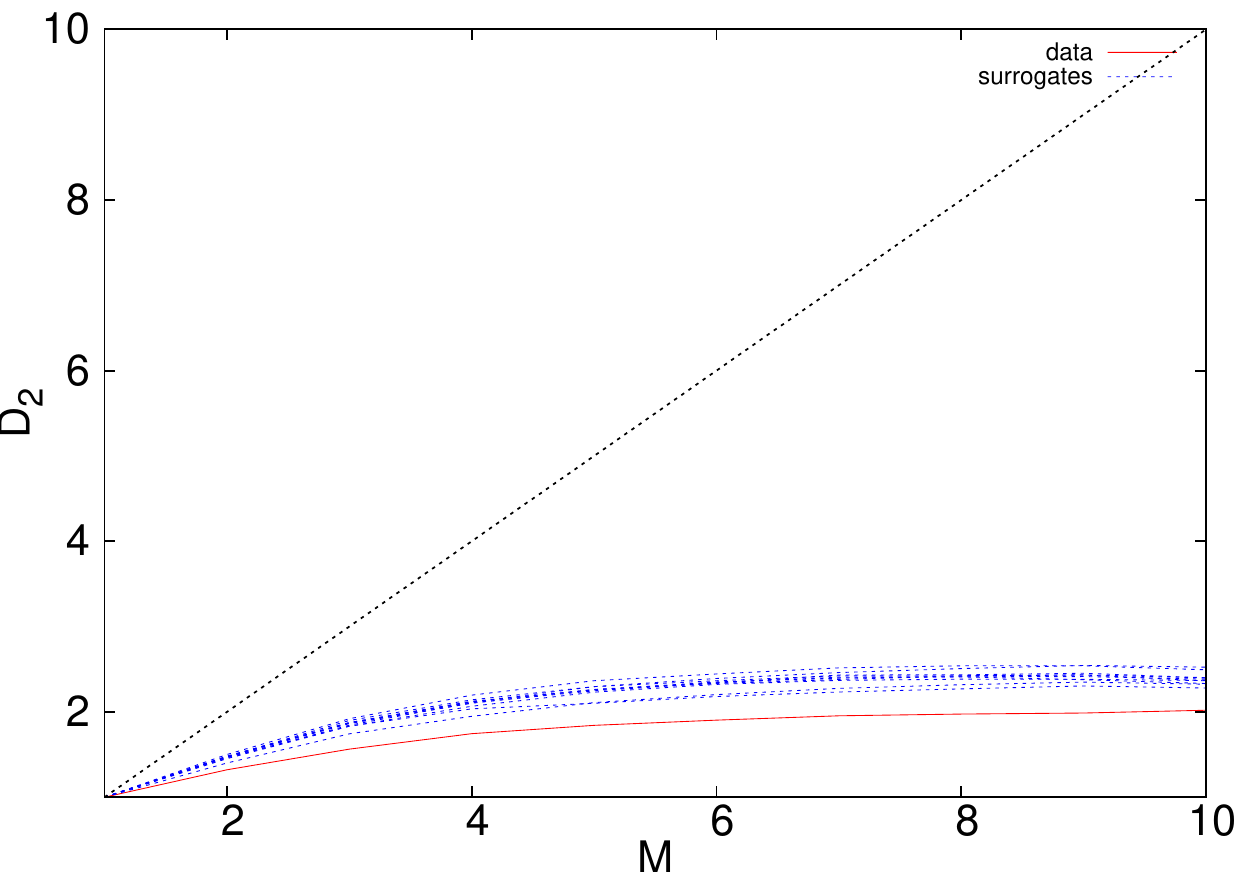}
\caption{Variation of correlation dimension $D_{2}$ as a function of embedding dimension $M$ for ADAF-like simulated lightcurves. The red-solid line represents the original data, the dashed-blue lines represent the surrogates while the straight diagonal black line represents the behavior of an ideal stochastic system.}
\label{fig:simd2}
\end{figure}

\subsubsection{Time-series analysis of simulated lightcurve}

Figure \ref{fig:simd2} shows the variation of $D_{2}$ with $M$ for a typical ADAF-like solution (with no cooling). As evident, $D_{2}$ saturates at a low value of $M$ which is in tandem with our argument for the expected behavior of such a system. More so, the surrogates  (see e.g Schreiber \& Schmitz 1996) are distinct from the original data with respect to their $D_{2}$ values. 
This confirms indeed that the time series of ADAF corresponds to low $D_2$, as argued in previous sections in order to interpret observed data. In future works, we intend to explore this further with many more simulated lightcurves of many classes, using better fine tuned set up and incorporating cooling at some point, which is beyond the scope of the present work.  


\section{Discussions and Conclusion}

An accretion flow around a compact object is a nonlinear general relativistic system involving magnetohydrodynamics. Earlier, at least some of the accretion flows were argued to be similar to a Lorenz system (Mukhopadhyay 2004; Misra et al. 2004, 2006; Karak, Dutta \& Mukhopadhyay 2010), but contaminated by Poisson noise. These results were however questioned by
Mannattil, Gupta \& Chakraborty (2016). Even if they are not noise contaminated chaos, definitely different lightcurves 
(of the same source, like, GRS~1915+105 or different sources) exhibit different 
correlation dimensions, which must be related to some physics. In the Lorenz system, 
there exists a parameter, called control parameter ($R$), which determines the underlying dynamics, whether the system is chaotic (with a low $D_2$) or random (with very high unsaturated $D_2$). Similarly, in accretion flow, we propose that the accretion rate must play the role of one of the control parameters, which determines the dynamics of the flow. The other control parameters could be, e.g., viscosity and energy of the flow, which may all not be independent.

Out of the modes of accretion flows, in order to explain our results, the major ones considered are (1) Keplerian disc which is a geometrically thin but optically thick flow (Shakura \& Sunyaev 1973), (2) slim disc which is a radiation trapped geometrically thicker flow than the Keplerian disc, retaining its optically thick properties intact (Muchotrzeb \& Paczynski 1982; Abramowicz et al. 1988), (3) ADAF which is a geometrically thick, self-similar, quasispherical, sub-Keplerian, optically thin, hot flow (Narayan \& Yi 1994), and (4) GAAF which is in general a geometrically thick and optically thin (but not necessarily as thin as ADAF) sub-Keplerian flow and need not possess insignificant cooling mechanism (Rajesh \& Mukhopadhyay 2010). The last two classes of solutions naturally possess strong radial advection. By correlating/combining nonlinear temporal and spectral behaviors of RXTE data, we posit that the flow switches from one accretion regime to another with the change of accretion rate. In fact, most of the accretion flows are found to be admixture of these modes. However, the most important possibility in our correlation for nonlinear and spectral behaviors lies in determining the effective number of degrees of freedom (measuring $D_2$) in various accretion modes, which in turn helps to construct the model flows correctly. For example, while ADAF, GAAF, thin/slim discs are in the literatures over the years, it is still not very clear the correct number of variables describing them. This is very important to correctly interpret data based on model parameters. This was never attempted earlier.

Earlier, Belloni et al. (2000) and Reig, Belloni \& van der Klis (2003) investigated the connection between spectral states: $A, B, C$ and the canonical temporal behaviors of GRS~1915+105. They found that transition among spectral states is possible for a given temporal class, e.g. $\lambda$, although the class $\chi$ was shown to be in $C$ state throughout. Our analyses have however captured an alternative and different picture.
We have shown, e.g., that a temporal class of RXTE observation may consist of an admixture of Keplerian and sub-Keplerian components, which leads overall to a D-P spectral state. Of course, some classes are observed to exhibit DD or a pure LH state,
which respectively are Keplerian dominated flow with a high accretion rate or pure sub-Keplerian flow with a very low accretion rate. Therefore, the admixture of Keplerian and sub-Keplerian components with a D-P state, mentioned above, have accretion rates between high and low. Further, this is the transition state between pure Keplerian and pure sub-Keplerian states, which is possible by increasing (decreasing) accretion rate from a sub-Keplerian (Keplerian) flow. This indicates that, while the works by Belloni and his collaborators emphasized individual spectral states appearing in temporal classes, we are more interested in nonlinear features in accretion flows generating the overall weighted average spectral nature. Our result is very useful for a concrete modeling of magnetohydrodynamics of flows.

An ADAF necessarily possesses sub-critical low mass accretion rate making it hot, radiatively inefficient, optically thin and self-similar. For GRS~1915+105, classes $\theta, \nu, \alpha$ belong to this regime showing low saturated $D_2$ values --- signature of self-similarity. Once the mass accretion rate increases, the various cooling processes start working randomly, converting it to GAAF, generating $\chi$ class which is of stochastic nature. With the further increase of mass accretion rate upto the order of Eddington limit, the flow starts exhibiting efficient cooling, bringing it to the optically thick regime which reveals coherent flares. As a result, the flow might 
convert to a geometrically thin flow including a significant Keplerian component with an insignificant advection. At this stage the flow would reveal a low $D_2$ again, as seen 
in the classes $\beta, \lambda, \kappa, \mu$. With the further increase of mass accretion rate, the flow would be radiation dominated and geometrically thicker, rendering a slim disc (but with a nonnegligible PL component, hence non-pure slim disc). This would be geometrically much thicker than a Keplerian flow and again might reveal stochastic fluctuations due to the presence of excessive matter and radiation (soft and hard) randomly in all directions, as seen in $\delta, \phi$, $\gamma$  classes. Note that, as discussed above, 
the spectra with similar contributions from diskbb and PL (D-P state), along with lower values of $D_2$ in the corresponding time series, 
seem to be corresponding to an intermediate flow between GAAF and Keplerian disc (but more towards to Keplerian) when their  SI is 
generally around $3$ or slightly 
more. However, similar spectra with a high $D_2$ would plausibly signify an intermediate flow between Keplerian disc 
and slim disc when SI is as high as $\sim 3.5$. This is because, a slim disc would have less interaction with any 
possible surrounding sub-Keplerian region as the entire quasi-spherical region is mostly radiation trapped, hence will have higher SI. 
However, the Keplerian disc would have a surrounding quasi-spherical sub-Keplerian region contributing PL with a lesser SI. 

The above heuristic argument is plausibly one of the cycles occurring in accretion flows, rendering transitions between different flow regimes. The same cycle will be traced inversely, if the accretion rate decreases from a high value to a low value. The same cycle could also be understood by means of the change of flow angular momentum per unit mass (high in Keplerian and low in sub-Keplerian flows) which is expected to be related to the flow viscosity (in general high with a high angular momentum flow and vice versa) as well (see, e.g., Rajesh \& Mukhopadhyay 2010). Now other black hole sources, including AGNs, should also be similarly studied in order to understand the correlation between nonlinear and spectral behaviors in them to further confirm our results. 
For the neutron star Sco~X-1, it is found from the ObsIDs analyzed here that sometimes the flow remains in a mode between Keplerian and slim discs only. Based on this, it is revealed that the time series of Sco~X-1 exhibits differently varying nonlinear temporal classes, as of GRS~1915+105.

\section*{Acknowledgements}

This work is partly supported by an ISRO project with Grant No. ISTC/PPH/BMP/0362.
The authors would like to thank Mariano Mendez for clarifying certain computations, Ramesh Narayan for discussion and, finally, A. R. Rao and Nilkanth Vagshette for helping to obtain some results and discussion. Moreover, helps with providing simulated lightcurves of accretion flows by Christopher Reynolds and Scott Noble independently are gratefully acknowledged (although finally those lightcurves could not be used for the present purpose). Thanks are also due to Prateek Sharma for introducing us with the simulations we performed. Finally,
we thank the referee for useful suggestions to improve the presentation of the work.

\section*{Some extra material}
For the spectral fitting of the GRS~1915+105 classes, we consider the energy range $3-25\,\mathrm{keV}$ and the Gaussian line energy is kept fixed at $6.4\,\mathrm{keV}$ in line with standard practice (e.g. Altamirano et al. 2011). We must add that while some of the data do not require a Gaussian for a good fit, others require more than one Gaussian line (possibly due to significant disc reflection). ObsIDs 10408-01-12-00 ($\phi$) and 20402-01-56-00 ($\gamma$) require an additional Gaussian line each (besides the iron line) at energies $4.4\,\mathrm{keV}$ and $5.3\,\mathrm{keV}$ respectively to give acceptable fits with physically plausible parameters. The standard absorption model  "wabs" is used to account for interstellar absorption and the column density $n_{H}$ is fixed at $7\times 10^{22}$ atoms~cm$^{-2}$. The best fit parameters are shown in Table \ref{tab:tab3}

\begin{table*}
\caption{GRS~1915+105: Model best fit parameters}
\label{tab:tab3}
Model: wabs~(powerlaw + diskbb): a Gaussian line is added where needed
{\small\small
\begin{tabular}{cccccccccccccccccccccc}\\
\hline
\hline
ObsID & class & $n_{PL}$ & $\sigma_{ga}$ & $n_{ga}$ & diskbb $T_{in}$ & $n_{disc}$ & $F~(10^{-8})$ \\
\hline
\hline
\hline
10408-01-10-00 & $\beta$  & $87.8^{+7.1}_{-7.1}$ & $0.79^{+0.17}_{-0.17}$ & $0.09^{+0.02}_{-0.02}$ & $1.98^{+0.02}_{-0.02}$ & $117.5^{+10.8}_{-10.8}$ & $3.7371$\\
\hline
20402-01-37-01 & $\lambda$ & $36.1^{+2.9}_{-2.9}$ & $-$ & $-$ & $2.09^{+0.02}_{-0.02}$ & $82.2^{+4.4}_{-4.4}$ & $2.9878$\\
\hline
20402-01-33-00 & $\kappa$ & $37.1^{+2.7}_{-2.7}$ & $-$ & $-$ & $2.08^{+0.02}_{-0.02}$ & $67.1^{+4.0}_{-4.0}$ & $2.6055$\\
\hline
10408-01-08-00 & $\mu$ &  $74.0^{+7.8}_{-7.8}$ & $0.91^{+0.15}_{-0.15}$ & $0.11^{+0.02}_{-0.02}$ & $2.00^{+0.02}_{-0.02}$ & $138.0^{+11.4}_{-11.4}$ & $3.8698$\\
\hline
20402-01-45-02 & $\theta$ &  $102.2^{+4.7}_{-4.7}$ & $0.10^{*}$ & $0.05^{+0.01}_{-0.01}$ & $1.98^{+0.09}_{-0.09}$ & $24.3^{+8.0}_{-8.0}$ & $3.3871$  \\
\hline
10408-01-40-00 & $\nu$ &  $59.5^{+4.0}_{-4.0}$ & $-$ & $-$ & $1.67^{+0.02}_{-0.02}$ & $149.5^{+15.0}_{-15.0}$ & $3.3505$\\
\hline
20187-02-01-00 & $\alpha$ &  $8.1^{+0.5}_{-0.5}$ & $-$ & $-$ & $1.55^{+0.02}_{-0.02}$ & $211.0^{+18.5}_{-18.5}$ & $1.2850$ \\
\hline
20402-01-03-00 & $\rho$ &  $39.0^{+2.8}_{-2.8}$ & $-$ & $-$ & $1.55^{+0.02}_{-0.02}$ & $169.9^{+14.8}_{-14.8}$ & $2.5473$\\
\hline
10408-01-17-00 & $\delta$ &  $80.8^{+6.6}_{-6.6}$ & $0.67^{+0.14}_{-0.14}$ & $0.07^{+0.01}_{-0.01}$ & $1.90^{+0.02}_{-0.02}$ & $101.7^{+9.1}_{-9.1}$ & $2.5697$\\
\hline
10408-01-12-00 & $\phi$ & $83.6^{+3.1}_{-3.1}$ & $0.86^{+0.05}_{-0.05}$ & $0.13^{+0.01}_{-0.01}$ & $2.07^{+0.02}_{-0.02}$ & $52.2^{+2.7}_{-2.7}$ & $1.9109$\\
\hline
20402-01-56-00 & $\gamma$ & $126.0^{+13.9}_{-13.9}$ & $0.80^{*}$ & $0.44^{+0.09}_{-0.09}$ & $2.19^{+0.02}_{-0.02}$ & $85.8^{+5.3}_{-5.3}$ & $3.4465$\\
\hline
10408-01-22-00 & $\chi$ & $50.5^{+1.0}_{-1.0}$ & $0.52^{+0.14}_{-0.14}$ & $0.04^{+0.01}_{-0.01}$ & $2.98^{*}$ & $1.9^{+0.2}_{-0.2}$ & $2.0180$\\
\hline
\hline
\hline
\end{tabular}
}


{Columns:- 
1: RXTE observational identification number.
2: Temporal class.
3: The powerlaw normalization.
4: Gaussian sigma (keV).
5: The normalization of the Gaussian line.
6: The disc blackbody temperature in keV.
7: The disc normalization.
8: Total flux in $3-25$ keV (ergs/cm$^2$/s). 
N.B: The parameters with '$*$' are kept fixed for acceptable spectral fits
}
\end{table*}



For the spectral fitting of Sco~X-1 the energy range considered is $3-25\,\mathrm{keV}$ and the Gaussian line energy is kept fixed at $6.4\,\mathrm{keV}$ (e.g. Bradshaw et al. 2003) where necessary, otherwise it is excluded. The standard absorption model "wabs" is used to account for interstellar absorption and the column density $n_{H}$ is fixed at $0.15\times 10^{22}$ atoms~cm$^{-2}$. The best fit parameters are given in Table \ref{tab:tab4}

\begin{table*}
\caption{Sco~X-1: Model best fit parameters}
\label{tab:tab4}
Model: wabs~(bb + diskbb): a Gaussian line is added where necessary
{\small
\begin{tabular}{cccccccccccccccccccccc}\\
\hline
\hline
ObsID & $\sigma_{ga}$ [keV] & $n_{ga}$ & $T_{bb}$ & $n_{bb}$ & diskbb $T_{in}$ & $n_{disc}$ & $F~(10^{-7})$ \\
\hline
\hline
91012-01-02-00  & $0.23^{+0.53}_{-0.53}$ & $0.10^{+0.05}_{-0.05}$ & $2.81^{+0.07}_{-0.07}$ & $0.57^{+0.09}_{-0.09}$ & $2.21^{+0.08}_{-0.08}$ & $407.0^{+45.4}_{-45.4}$ & $1.7635$\\
 70014-01-02-02  & $0.43^{+0.25}_{-0.25}$ & $0.17^{+0.05}_{-0.05}$ & $2.58^{+0.03}_{-0.03}$ & $0.70^{+0.04}_{-0.04}$ & $1.81^{+0.04}_{-0.04}$ & $832.8^{+72.1}_{-72.1}$ & $1.6134$\\
 40020-01-01-07  & $-$ & $-$ & $2.64^{+0.02}_{-0.02}$ & $0.91^{+0.02}_{-0.02}$ & $1.66^{+0.03}_{-0.03}$ & $1081.7^{+76.4}_{-76.4}$ & $1.6059$\\
\hline
\end{tabular}
}\\
{Columns:- 
1: RXTE observational identification number.
2: Gaussian line sigma in keV.
3: Gaussian line normalization.
4: Blackbody temperature in keV.
5: Blackbody normalization.
6: The disc blackbody temperature in keV.
7: The disc blackbody normalization.
8: Total flux in $3-25$ keV (ergs/cm$^2$/s).
}
\end{table*}


\bsp	
\label{lastpage}

\begin{thebibliography}{}

\bibitem[adegoke (2017)]{} Adegoke, O., Rakshit, S., \&  Mukhopadhyay B., \ 2017, MNRAS, 466, 3951
\bibitem[czerny (1988)]{} Abramowicz, M. A., Czerny, B., Lasota, J. P., Szuszkiewicz, E., \ 1988, ApJ, 332, 646
\bibitem[altamiramo (2011)]{} Altamirano, D., et al., \ 2011, ApJ, 742, 17
\bibitem[belloni (2005)]{} Belloni, T., \ 2005, Interacting Binaries: Accretion, Evolution and Outcomes, eds. L. A. Antonelli, et al., Procs. Interacting Binaries Meeting, Cefalu, Italy, June, \ 2004; astro-ph/0504185.
\bibitem[belloni (2000)]{} Belloni, T., Klein-Wolt, M., M\'endez, M., van der Klis, M., van Paradijs, J., \ 2000, A\&A, 355, 271
\bibitem[belloni (1997)]{} Belloni, T., M\'endez, M., King, A. R., van der Klis, M., van Paradijs, J., \ 1997, ApJ, 488, L109
\bibitem[belloni (2006)]{} Belloni, T., Soleri, P., Casella, P., M\'endez, M., Migliari, S., \ 2006, MNRAS, 369, 305
\bibitem[bradshaw (1999)]{} Bradshaw, C. F., Fomalont, E. B., \& Geldzahler, B. J., \ 1999, ApJ, 512, L121
\bibitem[bradshaw (2003)]{} Bradshaw, C. F., Geldzahler, B. J., \& Fomalont, E. B., \ 2003, ApJ, 592, 486
\bibitem[chakrabarti (1995)]{} Chakrabarti, S. K., \& Titarchuk, L. G., \ 1995, ApJ, 455, 623
\bibitem[das (2013)]{das_sharma} Das, U., Sharma, P., \ 2013, MNRAS, 435, 2431
\bibitem[Flock (2010)]{} Flock, M., Dzyurkevich, N., Klahr, H., Mignone, A., \ 2010 A\&A, 516, A26
\bibitem[fender (2004)]{} Fender, R., \& Belloni, T., \ 2004, Ann. Rev. Astron. Astrophys., 42, 317
\bibitem[gardiner (2005)]{gs2005} Gardiner, T. A., Stone, J. M., \ 2005, JCoPh, 205, 509
\bibitem[greiner (2001)]{} Greiner, J., Cuby, J. G., \& McCaughrean, M. J., \ 2001, Nature, 414, 522
\bibitem[grassberger-procacciar (1983)]{} Grassberger, P., Procaccia, I., \ 1983, Phy. D, 9, 189
\bibitem[guan (2009)]{Guan2009} Guan, X., Gammie, C. F., Simon, J. B., Johnson, B. M., \ 2009, ApJ, 694, 1010
\bibitem[hari (2009)]{} Harikrishnan, K. P., Misra, R., Ambika, G., Amritkar, R. E., \ 2009, Chaos, 19, 3129
\bibitem[hasinger (1989)]{} Hasinger, G., \& van der Klis, M., \ 1989, A\&A, 225, 79
\bibitem[hawley (2011)]{hawley_krolik2011} Hawley, J. F., Guan, X., Krolik, J. H., \ 2011, ApJ, 738, 84
\bibitem[homan (2005)]{} Homan, J., Belloni, T., \ 2005, Ap\&SS, 300, 107
\bibitem[ingram (2011)]{} Ingram, A., Done, C., \ 2011, MNRAS, 415, 2323
\bibitem[kalamkar (2015)]{} Kalamkar, M., Reynolds, M. T., van der Klis, M., Altamirano, D., Miller, J. M., \ 2015, ApJ, 802, 23
\bibitem[karak (2010)]{} Karak, B. B., Dutta, J., Mukhopadhyay, B., \ 2010, ApJ, 708, 862
\bibitem[klein (2002)]{} Klein-Wolt, M., Fender, R. P., Pooley, G. G., Belloni, T., Migliari, S., Morgan, E. H., van der Klis, M., \ 2002, MNRAS, 331, 745
\bibitem[kuulkers (1997)]{} Kuulkers, E., van der Klis, M., Oosterbroek, T., van Paradijs, J., \& Lewin, W. H. G., \ 1997, MNRAS, 287, 495
\bibitem[liang (1998)]{} Liang, P., \ 1998, Phys. Rep., 302, 67
\bibitem[mendez (1997)]{} Mendez, M., van der Klis, M., \ 1997, ApJ, 479, 926
\bibitem[mannattil (2016)]{sagar} Mannattil M., Gupta H., \& Chakraborty S., \ 2016, ApJ, 833, 208
\bibitem[migliari (2003)]{} Migliari, S., Belloni, T., \ 2003, A\&A, 404, 283
\bibitem[mignone (2007)]{mignone} Mignone, A., Bodo, G., Massaglia, S., Matsakos, T., Tesileanu, O., Zanni, C., Ferrari, A., \ 2007, ApJS, 170, 228
\bibitem[mirabel (1994)]{} Mirabel, I. F., Rodr\`iguez, L. F., \ 1994, Nature, 371, 46
\bibitem[misra (2006)]{}  Misra, R., Harikrishnan, K. P., Ambika, G., Kembhavi, A. K., \ 2006, ApJ, 643, 1114
\bibitem[misra (2004)]{}  Misra, R., Harikrishnan, K. P., Mukhopadhyay, B., Ambika, G., Kembhavi, A. K., \ 2004, ApJ, 609, 313
\bibitem[miyamoto (1992)]{} Miyamoto, S., Kitamoto, S., Iga, S., Negoro, H., Terada, K., \ 1992, ApJ, 391, L21
\bibitem[miyoshi (2005)]{} Miyoshi, T.; Kusano, K. 2005AGUFMSM51B1295M
\bibitem[muchotrzeb (1982)]{} Muchotrzeb, B., Paczynski, B., \ 1982, AcA, 32, 1
\bibitem[mukhopadhyay (2004)]{} Mukhopadhyay, B., \ 2004, AIPC, 714, 48
\bibitem[muno (2002)]{} Muno, M. P., Remillard, R.A., \& Chakrabarty, D., \ 2002, ApJ, 568, L35
\bibitem[narayan (1994)]{} Narayan, R., Yi, I., \ 1994, ApJ, 428, L13
\bibitem[narayan (1995)]{} Narayan, R., Yi, I., \ 1995, ApJ, 452, 379
\bibitem[noble (2010)]{Noble2010} Noble, S. C., Krolik, J. H., Hawley, J. F., \ 2010, ApJ, 711, 959
\bibitem[paczynsky (1980)]{pw} Paczynsky, B., Wiita, P. J., \ 1980, A\&A, 88, 23
\bibitem[papaloizou (1984)]{papa_pri} Papaloizou, J. C. B., Pringle, J. E., \ 1984, MNRAS, 208, 721
\bibitem[pessah (2010)]{Pessah2010} Pessah, M. E., \ 2010, ApJ, 716, 1012
\bibitem[pooley (1997)]{} Pooley, G. G., Fender, R. P., \ 1997, MNRAS, 292, 925
\bibitem[rajesh (2010)]{} Rajesh, S. R., Mukhopadhyay, B., \ 2010, MNRAS, 402, 961
\bibitem[reig (2003)]{} Reig, P., Belloni, T., van der Klis, M., \ 2003, A\&A, 412, 229
\bibitem[remillard (2006)]{} Remillard, R. A., McClintock, J. E., \ 2006, ARA\&A, 44, 49
\bibitem[schreiber (1996)]{} Schreiber T \& Schmitz A., \ 1996, Physical Review Letters, 77, 635
\bibitem[shakura (1973)]{} Shakura, N. I., Sunyaev, R. A., \ 1973, A\&A, 24, 337
\bibitem[sorathia (2012)]{sorathia2012} Sorathia, K. A., Reynolds, C. S., Stone, J. M., Beckwith, K., \ 2012, ApJ, 749, 189
\bibitem[steeghs (2002)]{} Steeghs, D., \& Casares, J., \ 2002, ApJ, 568, 273
\bibitem[szuszkiewicz (1996)]{} Szuszkiewicz, E., Malkan, M. A., \& Abramowicz, M. A., \ 1996, ApJ, 458, 474
\bibitem[tanaka (1996)]{} Tanaka, Y., Shibazaki, N., \ 1996, ARA\&A, 34, 607
\bibitem[tananbaum (1972)]{} Tananbaum, H., Gursky, H., Kellogg, E., Giacconi, R., Jones, C., \ 1972, ApJ, 177, L5
\bibitem[van (1995)]{} van der Klis, M., \ 1995, in X-ray binaries, ed. W. H. G. Lewin, J. van Paradijs, \& E. P. J. van den Heuvel (Cambridge University Press), 252
\bibitem[zdziarski (2004)]{} Zdziarski, A. A., Gierlinski, M., \ 2004, PThPS, 155, 99

\end{thebibliography}
\end{document}